\pdfoutput=1
\documentclass[a4paper,american,floatfix,pdftex,superscriptaddress,reprint,onecolumn,secnumarabic,notitlepage]{revtex4-1}

\usepackage{natbib}

\usepackage{dcolumn}
\usepackage{graphicx}
\usepackage{url}
\usepackage{color}
\usepackage{url}
\usepackage{verbatim}

\newcommand{\cm}{cm$^{-1}$}

\newcommand{\highlight}[1]{}

\newcommand{\Duo}{{\sc Duo}}
\newcommand{\Level}{{\sc Level}}

\newcommand{\PGOPHER}{{\sc PGOPHER}}
\newcommand{\RKR}{{\sc RKR1}}

\newcommand{\midrule}{\hline\hline}

\newcommand{\bottomrule}{\hline\hline}

\begin{document}

\title{Empirical Line Lists in the ExoMol Database}

\author{Yixin Wang}
\affiliation{Department of Physics and Astronomy, University College London, Gower Street, WC1E 6BT London, United Kingdom}

\author{Jonathan Tennyson\footnote{Corresponding author: j.tennyson@ucl.ac.uk}}
\affiliation{Department of Physics and Astronomy, University College London, Gower Street, WC1E 6BT London, United Kingdom}

\author{Sergei N. Yurchenko}
\affiliation{Department of Physics and Astronomy, University College London, Gower Street, WC1E 6BT London, United Kingdom}

\begin{abstract}
The ExoMol database aims to provide comprehensive molecular
line lists for exoplanetary and other hot atmospheres. The data are expanded by
inclusion of empirically derived line lists taken from the literature for a series of diatomic molecules,
namely CH, NH, OH, AlCl, AlF, OH$^+$, CaF, MgF, KF, NaF, LiCl, LiF, MgH, TiH, CrH,
FeH,  C$_2$, CP, CN, CaH, and triplet N$_2$. Generally, these line lists are
constructed from measured spectra using a combination of effective rotational Hamiltonian models for the line positions and ab initio (transition) dipole moments to provide intensities. This~work results in the inclusion of 22 new
molecules (36 new isotopologues) in the ExoMol database.
\end{abstract}

\maketitle

\section{Introduction}

The ExoMol project \citep{jt528} aims to provide comprehensive spectroscopic data for molecules which are important for exoplanetary and other hot astronomical atmospheres. Thus far, these data have been provided in the form of computed line lists generated as part of the ExoMol project itself or taken from other sources which use a similar methodology. In~principle, it is possible to compute line lists by a direct numerical
solution of the appropriate Schr\"odinger equation and the procedure for doing this, which
involves initially using the Born--Oppenheimer approximation to separate the electronic
and nuclear motion, is now well-established \citep{jt475}. However, while there are
a number of codes available \citep{jt626} which can provide accurate solutions to the few-body nuclear motion problem using the variational principle, it is generally not
possible to solve the electronic structure problem with the required accuracy for most molecules of interest; see~\cite{jt623} for a discussion of some of the issues
involved.

The ExoMol project has therefore used a methodology \citep{jt511,jt693} which involves the construction of an appropriate spectroscopic model based on a combination of ab~initio potential energy curves (PECs), dipole moment functions (DMFs) and, where appropriate, coupling curves. It is found that, with the use of appropriate electronic structure methods,
ab~initio dipole moment surfaces can provide reliable transition intensities \citep{jt573} even with quantified uncertainties if these are needed \citep{jt625}.
However, to obtain accurate transition frequencies, the ab~initio potential energy surfaces (or curves for diatomic molecules) need to be refined using experimental data;
see~\cite{jt503} for a discussion of the procedure used. For~open shell systems,
various angular momenta, spin-orbit, spin-rotation, etc. interactions can couple the potential energy curves; see~\cite{jt632}.
These couplings may also be refined.
We have developed a suite of programs for this task \citep{jt609,jt626}
based on a variational solution of the nuclear Schr{\"o}dinger equation.
The final problem solved using these programs, and~the programs themselves, can
be referred to as the spectroscopic model used to characterize the particular line list.
Provision of this model ensures not only that the results are reproducible, but also
allows results to be improved or extended using the model as a starting point.
Indeed, post hoc improvement of our line lists is an important part of our
general procedure; see~\cite{jt570}, for~example.

The current data base (\url{www.exomol.com} \citep{jt631}) contains line lists for some 60 molecules generated in this fashion. In~many cases, there
are corresponding line lists for isotopically substituted species. A~number of the ExoMol line lists contain many billions of~transitions.

However, line lists can be generated directly using laboratory measurements
\citep{09Bernath.exo}.
While these line lists are rarely as complete (or as large) as computed ones,
they can be significantly more accurate, particularly in the line positions.
These measurements can also provide data on systems for which computed line lists are presently unavailable, or~indeed, in some cases, they would be very difficult to construct. It is therefore useful to augment the current ExoMol database with appropriate
empirical line lists. The~purpose of this article is to report progress in doing just~that.

In practice, the~methodology used to construct the computed and empirical line lists can be closer than is apparent at first sight. While computed line lists often use experimental data to improve not only the potentials used but also the positions of both whole vibrational bands \citep{jt500} and individual transitions (eg \citet{jt570}), as~discussed below, the~empirical line lists often use dipole moment curves
computed ab~initio to generate line intensities. In~this case, the
final line lists may be generated either via effective Hamiltonian fits to the laboratory data, using programs such as \PGOPHER\ \citep{PGOPHER}, or~by
direct solution of the appropriate nuclear motion Schr\"odinger equation, for~which Le Roy's program  \Level\ \citep{LEVEL} is often used. Thus, the empirical
approach considered here, see~\cite{MOLLIST}, also
provides a spectroscopic model for each line~list.


\section{Method}

The ExoMol format of line lists assumes two files, a~States file (.states) and a Transitions file (.trans). There is one such file for each isotopologue considered. The~states file contains term energies (\cm) and
the state
description. This description consists of the state's IDs, total statistical weights, total rotational angular momentum  $J$  (integer or half-integer), and other quantum numbers (vibrational, rotational, electronic, symmetry, etc.), state lifetimes, and Land\'{e} $g$-factors \citep{jt656}. An~extension to include the uncertainty in
the energy will be implemented in the next release of the data base \citep{jt799}.
The~transition file contains the upper and lower state IDs, Einstein~$A$ coefficients (s$^{-1}$) and, optionally, transition frequencies (\cm). As~an example, Tables~\ref{t:states} and \ref{t:trans} give extracts from
ExoMol States and Transition files for the new $^{7}$Li$^{35}$Cl line~list.

Most of the experimental line lists presented in this work were collected from the MoLLIST website \url{http://bernath.uwaterloo.ca/molecularlists.php} \citep{MOLLIST},
the exception being the N$_2$ line list which is due to \citet{18WeCaCr.N2}.
These line lists were then converted to the ExoMol format \citep{jt631} using the following standard steps: (i) Extract states which correspond to unique sets of quantum numbers and energies and convert into the ExoMol format: all states are labelled with a unique ID which is simply the counting number of the state in the .states file (line number); (ii) Using the state IDs, generate a .trans file in the ExoMol format: ID$_f$, ID$_i$, $A_{fi}$ and $\tilde{\nu}_{fi}$. This procedure was used for all species except N$_2$, NH and CH. In~case of N$_2$, the~ExoMol format was generated directly from the latest version of \PGOPHER\ \citep{18WeCaCr.N2} using this new \PGOPHER\ feature. For C$_2$, CH and NH, the line lists were reformatted by \citet{18ViSmPr}.

The partition functions were computed using (in most cases) the data by \citet{84SaTaxx.partfunc} or \citet{16BaCoxx.partfunc}, and~corrected to
be in the ``physicists'' convention \citep{jt777}, as~used by ExoMol and HITRAN \citep{jt692}; this convention includes the full nuclear spin degeneracy.
The lifetimes were generated using the methodology reported by \citet{jt624}.

\begin{table}[h!]
\centering
\caption{Extract from the states file of the $^{7}$Li$^{35}$Cl line~list. }
{

\label{t:states}
{\begin{tabular}{rrrrrrcclrrrr} \toprule
\boldmath$i$ & \textbf{Energy (\cm)} & \boldmath$g_i$ & \boldmath$J$ & \boldmath$\tau$ & \boldmath$v$	\\
\midrule
           1  &   0.000000 &    16  &     0 & {-1.0000E+00}  &    0 \\ 
           2  &   1.405000 &    48  &     1 &  {6.4935E+04}  &    0 \\
           3  &   4.215000 &    80  &     2 &  {6.8027E+03}  &    0 \\
           4  &   8.430000 &   112  &     3 &  1.8762E+03  &    0 \\
           5  &  14.049000 &   144  &     4 &  7.6336E+02  &    0 \\
           6  &  21.072000 &   176  &     5 &  3.8168E+02  &    0 \\
           7  &  29.499000 &   208  &     6 &  2.1786E+02  &    0 \\
           8  &  39.331000 &   240  &     7 &  1.3569E+02  &    0 \\
           9  &  50.564000 &   272  &     8 &  9.0090E+01  &    0 \\
          10  &  63.200000 &   304  &     9 &  6.2893E+01  &    0 \\
          11  &  77.237000 &   336  &    10 &  4.5662E+01  &    0 \\
          12  &  92.674000 &   368  &    11 &  3.4130E+01  &    0 \\
          13  & 109.511000 &   400  &    12 &  2.6247E+01  &    0 \\
          14  & 127.747000 &   432  &    13 &  2.0576E+01  &    0 \\
          15  & 147.380000 &   464  &    14 &  1.6447E+01  &    0 \\
\bottomrule
\end{tabular}}
\begin{tabular}{@{}c@{}}
\multicolumn{1}{p{\textwidth -.88in}}{\footnotesize $i$:   State counting number; $\tilde{E}$: State energy in \cm; $g_i$:  Total statistical weight, equal to ${g_{\rm ns}(2J + 1)}$; $J$: Total angular momentum; $\tau$: Lifetime (s$^{-1}$); $v$:   State vibrational quantum number.}
\end{tabular}

}

\end{table}
\unskip

\begin{table}[h!]
\centering
\caption{Extract from the transitions file of the $^{7}$Li$^{35}$Cl line~list.}
\label{t:trans}
\centering
\begin{tabular}{rrrr} \toprule
\multicolumn{1}{c}{\boldmath$f$}	&	\multicolumn{1}{c}{\boldmath$i$}	& \multicolumn{1}{c}{\boldmath$A_{fi}$ \textbf{(s\boldmath$^{-1}$)}}	&\multicolumn{1}{c}{\boldmath$\tilde{\nu}_{fi}$} \\ \midrule
         492   &       475& {1.1900E+01}&     110.073000 \\
        1381   &      1370& {1.5400E+01}&     110.207000 \\
         951   &       939& {1.3300E+01}&     110.289000 \\
        1190   &      1178& 1.4400E+01&     110.386000 \\
         859   &       840& 1.3100E+01&     110.544000 \\
        1323   &      1313& 1.5200E+01&     110.605000 \\
        1122   &      1110& 1.4200E+01&     110.716000 \\
         738   &       721& 1.2800E+01&     110.780000 \\
        1264   &      1252& 1.5000E+01&     110.988000 \\
         624   &       606& 1.2600E+01&     110.997000 \\
        1049   &      1035& 1.3900E+01&     111.028000 \\
        1390   &      1381& 1.5900E+01&     111.110000 \\
         513   &       492& 1.2300E+01&     111.193000 \\
\bottomrule
\end{tabular} \\
\begin{tabular}{@{}c@{}}
\multicolumn{1}{p{\textwidth -.88in}}{\footnotesize $f$: Upper  state counting number; $i$:  Lower  state counting number; $A_{fi}$:  Einstein-$A$ coefficient in s$^{-1}$; $\tilde{\nu}_{fi}$: transition wavenumber in \cm.}
\end{tabular}

\end{table}

A summary of the empirical line lists (mostly from Bernath's group) is given in Table~\ref{t:linelists}.

\begin{table}[h!]
\caption{Empirical line list included in the ExoMol database. The~number of lines and number of states included in each line list are given. The~maximal temperatures of the partition functions is $T = 5000$~K. All MoLLIST empirical line lists are named `MoLLIST'. The~N$_2$ line list by \citet{18WeCaCr.N2}  is called `WCCRMT'. }

\label{t:linelists}
\centering
\scalebox{0.85}[0.85]{
\begin{tabular}{llrrrrl}
\toprule
\textbf{Molecule} &  \textbf{El. States} & \boldmath$v_{\rm max}$ & \boldmath$J^{\rm max}$ & \textbf{Lines} & \boldmath$N$ \textbf{States} &  \textbf{Ref.} \\
\midrule
Al$^{35}$Cl	&	$X$~$^1\Sigma^+$	&	11	&	200	&	{20245}
&	2423	&	\citet{18YoBexx.AlF} \\
Al$^{37}$Cl	&	$X$~$^1\Sigma^+$	&	11	&	200	&	{20245}	&	2423	&	\citet{18YoBexx.AlF} \\
AlF	&	$X$~$^1\Sigma^+$	&	11	&	200	&	40490	&	2423	&	\citet{18YoBexx.AlF} \\
$^{12}$CH	&	X~$^2\Pi$,  A~$^2\Delta$,  B~$^2\Sigma^-$, C~$^2\Sigma^+$	&	6	&	49.5	&	53,079	&	2526& \citet{14MaPlVa.CH} \\
$^{13}$CH	&	X~$^2\Pi$,  A~$^2\Delta$,  B~$^2\Sigma^-$, C~$^2\Sigma^+$	&	6	&	49.5	&	51,349	&	2428& \citet{14MaPlVa.CH} \\
NH	&	X~$^3\Sigma^-$, A~$^3\Pi$ &	6	&	45	&	22,545	&	1285	&	\citet{14BrBeWe.NH,15BrBeWe.NH,18FeBeHo.NH} \\
OH       & $X$~$^2\Pi$, $A$~$^2\Sigma^+$ & 13 & 58.5 & 54,276 & 1878 & \citet{16BrBeWe.OH,18YoBeHo.OH} \\
OH$^+$	&	X~$^3\Sigma^-$, A~$^3\Pi$	&	4	&	30	&	12,044	&	823	&	\citet{17HoBexx.OH+,18HoBiBe.OH+} \\
$^{40}$CaF	&	$X$~$^2\Sigma^+$	&	10	&	123	&	14,817	&	1363	&	\citet{18HoBexx.CaF} \\
$^{42}$CaF	&	$X$~$^2\Sigma^+$	&	10	&	123	&	14,817	&	1363	&	\citet{18HoBexx.CaF} \\
$^{43}$CaF	&	$X$~$^2\Sigma^+$	&	10	&	123	&	14,817	&	1363	&	\citet{18HoBexx.CaF} \\
$^{44}$CaF	&	$X$~$^2\Sigma^+$	&	10	&	123	&	14,817	&	1363	&	\citet{18HoBexx.CaF} \\
$^{46}$CaF	&	$X$~$^2\Sigma^+$	&	10	&	123	&	14,817	&	1363	&	\citet{18HoBexx.CaF} \\
$^{48}$CaF	&	$X$~$^2\Sigma^+$	&	10	&	123	&	14,817	&	1363	&	\citet{18HoBexx.CaF} \\
$^{39}$KF	&	$X$~$^1\Sigma^+$	&	10	&	110	&	10,572	&	1065	&	\citet{16FrBeBr.NaF} \\
$^{41}$KF	&	$X$~$^1\Sigma^+$	&	10	&	109	&	10,379	&	1047	&	\citet{16FrBeBr.NaF} \\
NaF	&	$X$~$^1\Sigma^+$	&	10	&	90	&	7884	&	839	&	\citet{16FrBeBr.NaF} \\
$^{24}$MgF	&	$X$~$^2\Sigma^+$	&	8	&	101	&	8136	&	917	&	\citet{17HoBexx.MgF} \\
$^{25}$MgF	&	$X$~$^2\Sigma^+$	&	8	&	101	&	8136	&	917	&	\citet{17HoBexx.MgF} \\
$^{26}$MgF	&	$X$~$^2\Sigma^+$	&	8	&	101	&	8136	&	917	&	\citet{17HoBexx.MgF} \\
$^6$Li$^{35}$Cl	&	$X$~$^1\Sigma^+$	&	11	&	201	&	26,260	&	2423	&	\citet{18BiBexx.LiF} \\
$^6$Li$^{37}$Cl	&	$X$~$^1\Sigma^+$	&	11	&	201	&	26,260	&	2423	&	\citet{18BiBexx.LiF} \\
$^7$Li$^{35}$Cl	&	$X$~$^1\Sigma^+$	&	11	&	201	&	26,260	&	2423	&	\citet{18BiBexx.LiF} \\
$^7$Li$^{37}$Cl	&	$X$~$^1\Sigma^+$	&	11	&	201	&	26,260	&	2423	&	\citet{18BiBexx.LiF} \\
$^6$LiF	&	$X$~$^1\Sigma^+$	&	11	&	201	&	10,621	&	2423	&	\citet{18BiBexx.LiF} \\
$^7$LiF	&	$X$~$^1\Sigma^+$	&	11	&	201	&	10,621	&	2423	&	\citet{18BiBexx.LiF} \\
MgH	&	$X$~$^2\Sigma^+$, $A$~$^2\Pi$, $B^\prime$~$^2\Sigma^+$,	&	11	&	50.5	&	30896	&	1935	&	\citet{13GhShBe.MgH} \\
TiH	&	$X$~$^4\Phi$, $A$~$^4\Phi$, $B$~$^4\Gamma$	&	5	&	50.5	&	199,072	&	5788	&	\citet{05BuDuBa.TiH} \\
CrH	&	$X$~$^6\Sigma^+$, $A$~$^6\Sigma^+$	&	3	&	39.5	&	13,824	&	1646	&	\citet{06ChMeRi.CrH} \\
FeH	&	$X$~$^4\Delta$, $F$~$^4\Delta$	&	4	&	50.5	&	93,040	&	3564	&	\citet{10WEReSe.FeH} \\
C$_2$	&	$a$~$^3\Pi_g$, $d$~$^3\Pi_u$	&	3	&	97	&	155,110	&	4653	&	\citet{13BrBeSc.C2} \\
CP	&	$X$~$^2\Sigma^+$, $A$~$^2\Pi$	&	8	&	55.5	&	28,752	&	2114	&	\citet{14RaBrWe.CP} \\
CN	&	$X$~$^2\Sigma^+$, $A$~$^2\Pi$, $B$~$^2\Sigma^+$	&	22	&	115.5	&	195,120	&	7703	&	\citet{14BrRaWe.CN} \\
CaH	&	$X$~$^2\Sigma^+$, $A$~$^2\Pi$, $B$~$^2\Sigma^+$, $E$~$^2\Pi$	&	4	&	50.5	&	19,095	&	1771	&	\citet{12LiHaRa.CaH,13ShRaBe.CaH} \\
N$_2$	&	$A$~$^3\Sigma^+_u$, $B$~$^3\Pi_g$, $B'$~$^3\Sigma^-_u$, $W^3\Delta_u$	&	29	&	75	&	7,182,000	&	40,380	&	\citet{18WeCaCr.N2} \\
\bottomrule
\end{tabular}}
\end{table}

\subsection{AlCl and~AlF}

Vibrational-rotational line lists for AlCl and AlF in their ground ($X$~$^1\Sigma^+$) electronic states  are provided by \citet{18YoBexx.AlF}. The~line lists were generated using experimental  high-resolution ro-vibrational and rotational line positions: 1100 AlF transitions by \citet{95ZhGuBr.AlF} and 1544 transitions  of Al$^{35}$Cl and Al$^{37}$Cl  by \citet{93HeDuBe.AlCl} were used to obtain empirically the potential energy curves (PECs) by fitting.  The~PECs were used to calculate ro-vibrational energy levels for the vibrational excitations $v=0$ to $v=11$ and up to $J_{max}=200$ using program \Level\ \citep{LEVEL}. Ab initio dipole moment functions covering the PECs turning points were calculated and used to determine line~intensities.

\begin{figure}[h!]
\centering
\includegraphics[width=0.8\textwidth]{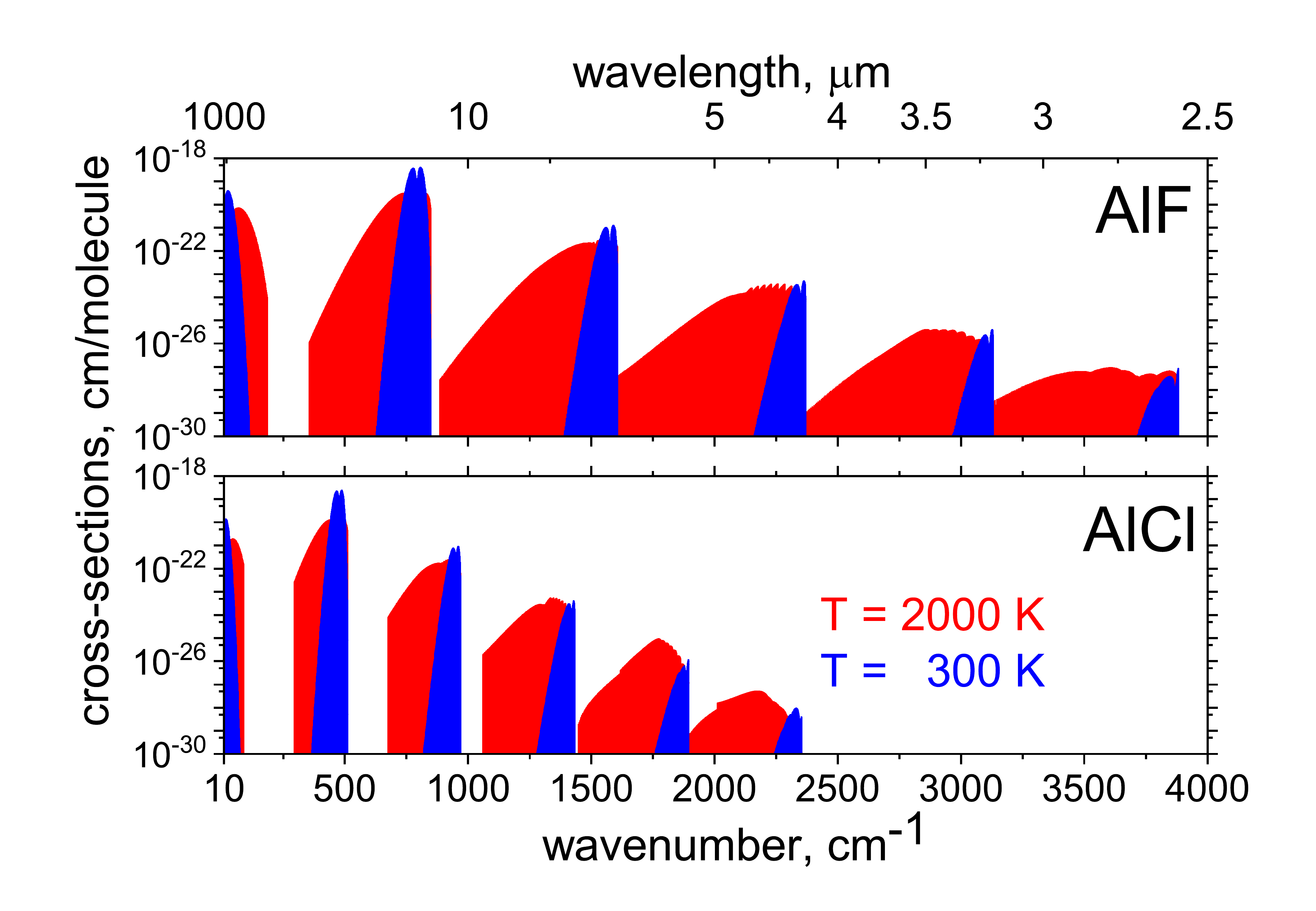}
\caption{Cross sections of AlCl and AlF computed using the line list  by \citet{18YoBexx.AlF} represented in the ExoMol~format.}
\label{f:AlF}
\end{figure}

Line lists were converted to ExoMol states and transition files; partition functions up to 5000~K were calculated with the program Exocross  \citep{jt708} using the energies from the line list. The~quantum numbers used for these line lists are simply $J$  and $v$.  Sample spectra are shown in Figure~\ref{f:AlF}.

\subsection{CH}

\begin{figure}[h!]
\includegraphics[width=0.8\textwidth]{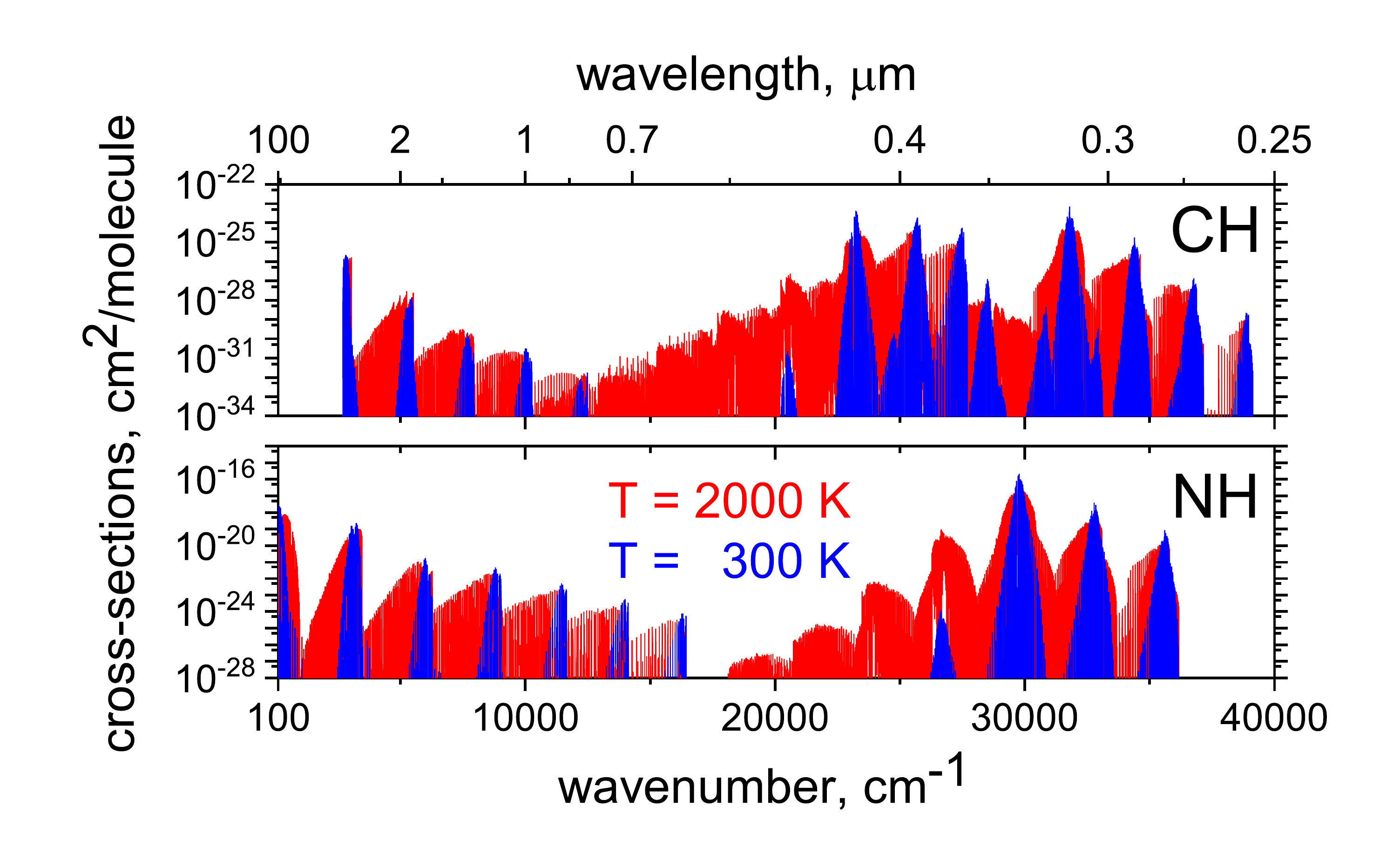}
\caption{Cross sections of {CN and NH}
computed using line lists by  \citep{14MaPlVa.CH, 14BrBeWe.NH,15BrBeWe.NH,18FeBeHo.NH}.
}
\label{f:CH:NH:OH}
\end{figure}

The original empirical line list for CH was produced by \citet{14MaPlVa.CH}. First, \PGOPHER\  was used to determine spectroscopic constants by fitting to the experimental data. The~global fit included the following experimental data: the X~$^2\Pi$ ro-vibrational fundamental transitions of \citet{10CoBexx.CH} from the ACE (Atmospheric Chemical Experiment); the solar spectrum of \citet{05BeMcAb.CH}; $A\,^2\Delta - X\,^2\Pi$ transitions  from the laboratory measurements by \citet{91BeBrOl.CH} and \citet{95Zachwieja.CH};  $B\,^2\Sigma^--X\,^2\Pi$ transitions due to \citet{96KePaRy.CH}, \citet{98KuHsHu.CH}, and~\citet{91BeBrOl.CH}; the $C\,^2\Sigma^+-X\,^2\Pi$ transitions by \citet{97BeKeRy.CH}, \citet{86UbMeMe.CH}, and~\citet{32Hexxxx.CH} for the 0--0 transition, \citet{99LiKuHs.CH} for the 1--1 transition, and~\citet{69HeJoxx.CH} for the 2--2 transition. Second-order Dunham constants were derived from the spectroscopic constants produced by \PGOPHER, and~then the Rydberg--Klein--Rees (RKR)
method, as~implemented in code \RKR\ \citep{RKR1}, was used to calculate the PECs. These PECs were used in \Level\  to generate the  transition moment matrix elements of $R(0)$, which were then used as input to \PGOPHER. In~a third step, a~CH line list was generated using \PGOPHER. Finally, this line lists were converted to ExoMol format by \citet{18ViSmPr},
and partition functions were calculated using the coefficients by \citet{84SaTaxx.partfunc}. An~important part of the conversion to the ExoMol format is to generate a unique set of energies. It is common for experimental line lists in the HITRAN format to have inconsistent energies representing different transitions, including the empirical line list by \citet{14MaPlVa.CH}. To provide a unique set of energy levels, such instances were replaced by the corresponding mean values. The~quantum numbers used are $J$, $v$, $e/f$ (rotationless parity), $N$ (rotational quantum number), and the electronic state labels X~$^2\Pi$, A~$^2\Delta$, B~$^2\Sigma^-$ and C~$^2\Sigma^+$. Illustrated spectra
generated using the line list are given in Figure~\ref{f:CH:NH:OH}.

\unskip

\subsection{NH}

The NH empirical line list consists of two parts taken from the work of \citet{14BrBeWe.NH,15BrBeWe.NH}, and \citet{18FeBeHo.NH}.
The~X~$^3\Sigma^-$ ground state line list by \citet{14BrBeWe.NH} contains ro-vibrational and rotational transitions, including fine structure, for~the vibrational bands up to $v'=6$ with $J_{\rm max}$ ranging from 25 to 45 for different vibrational bands and is based on the Einstein $A$ and $f$-values computed using an ab initio DMF which was generated with the internally contracted multi-reference configuration interaction (ic-MRCI) method and an aug-cc-pV6Z basis set. Programs \RKR, \Level, and~\PGOPHER\  were used to calculate line positions and intensities.  \citet{15BrBeWe.NH} used an
improved method to convert the transition matrix elements from Hund's case (b) to produce more accurate NH transition strengths. A~new line list for the $A\,^3\Pi-X\,^3\Sigma^-$ electronic band of NH was provided by \citet{18FeBeHo.NH}. The~spectroscopic constants for A~$^3\Pi$ and X~$^3\Sigma^-$ states were obtained from \citet{10RaBexx.NH} based on \citet{86BrRaBe.NH} for the $A$--$X$ transition and \citet{99RaBeHi.NH} for the infrared vibration--rotation and pure rotation lines. High level ab initio calculations were performed with MOLPRO  \citep{MOLPRO} to obtain the $A$--$X$ transition DMF. PECs and line strengths were calculated with Le Roy's \RKR\ and \Level\ programs, respectively. Line intensities and Einstein $A$ values were computed with  \PGOPHER\ after converting the Hund's case (b) output of \Level\ to the Hund's case (a) input needed for \PGOPHER.  Finally, the~line lists were compiled into the ExoMol format  by \citet{18ViSmPr} and the partition functions were calculated following \citet{84SaTaxx.partfunc}. See Table~\ref{t:gns} for the statistical weights used to convert their partition function to the physicists convention used
by ExoMol. Quantum numbers used are $J, v$, electronic state, $F$, $e/f$ and $N$. The~line list is illustrated in Figure~\ref{f:CH:NH:OH}.

\begin{table}[h!]
\caption{{Statistical} weights used to convert the partition functions by \citet{84SaTaxx.partfunc} to~ExoMol.}
\centering
\begin{tabular}{lrlr}
\toprule
\textbf{Molecule} & \boldmath$g_{\rm ns}$ & \textbf{Molecule} & \boldmath$g_{\rm ns}$ \\
\midrule
$^{27}$Al$^{35}$Cl &24 & $^{24}$Mg$^{19}$F & 2    \\
$^{27}$Al$^{37}$Cl &24 & $^{25}$Mg$^{19}$F & 12\\
$^{27}$Al$^{19}$F  & 12& $^{26}$Mg$^{19}$F & 2\\
$^{40}$Ca$^{19}$F  & 2 & $^{23}$Na$^{19}$F & 8\\
$^{42}$Ca$^{19}$F  & 2 & $^{14}$N$^{ 1}$H & 6\\
$^{43}$Ca$^{19}$F  & 16& $^{16}$O$^{ 1}$H & 2\\
$^{44}$Ca$^{19}$F  & 2 & $^{16}$O$^{ 1}$H+ & 2\\
$^{46}$Ca$^{19}$F  & 2 & $^{12}$C$^1$H & 2\\
$^{ 8}$Ca$^{19}$F  & 2 & $^{24}$Mg$^1$H & 2\\
$^{39}$K$^{19}$F   & 8 & $^{48}$Ti$^1$H & 2\\
$^{41}$K$^{19}$F   & 8 & $^{52}$Cr$^1$H & 2\\
$^{ 6}$Li$^{35}$Cl & 12& $^{56}$Fe$^1$H & 2\\
$^{ 6}$Li$^{37}$Cl & 12& $^{12}$C$_2$ & 1\\
$^{ 7}$Li$^{35}$Cl & 16&$^{12}$C$^{31}$P & 2\\
$^{ 7}$Li$^{37}$Cl & 16&$^{12}$C$^{14}$N & 3\\
$^{ 6}$Li$^{19}$F  & 6 &$^{40}$Ca$^{1}$H & 2\\
$^{ 7}$Li$^{19}$F  & 8 &
$^{14}$N$_2$ & 3 \& 6 \\
\bottomrule
\end{tabular}
\label{t:gns}
\end{table}
\unskip

\subsection{OH}

\begin{figure}[h!]
\centering
\includegraphics[width=0.8\textwidth]{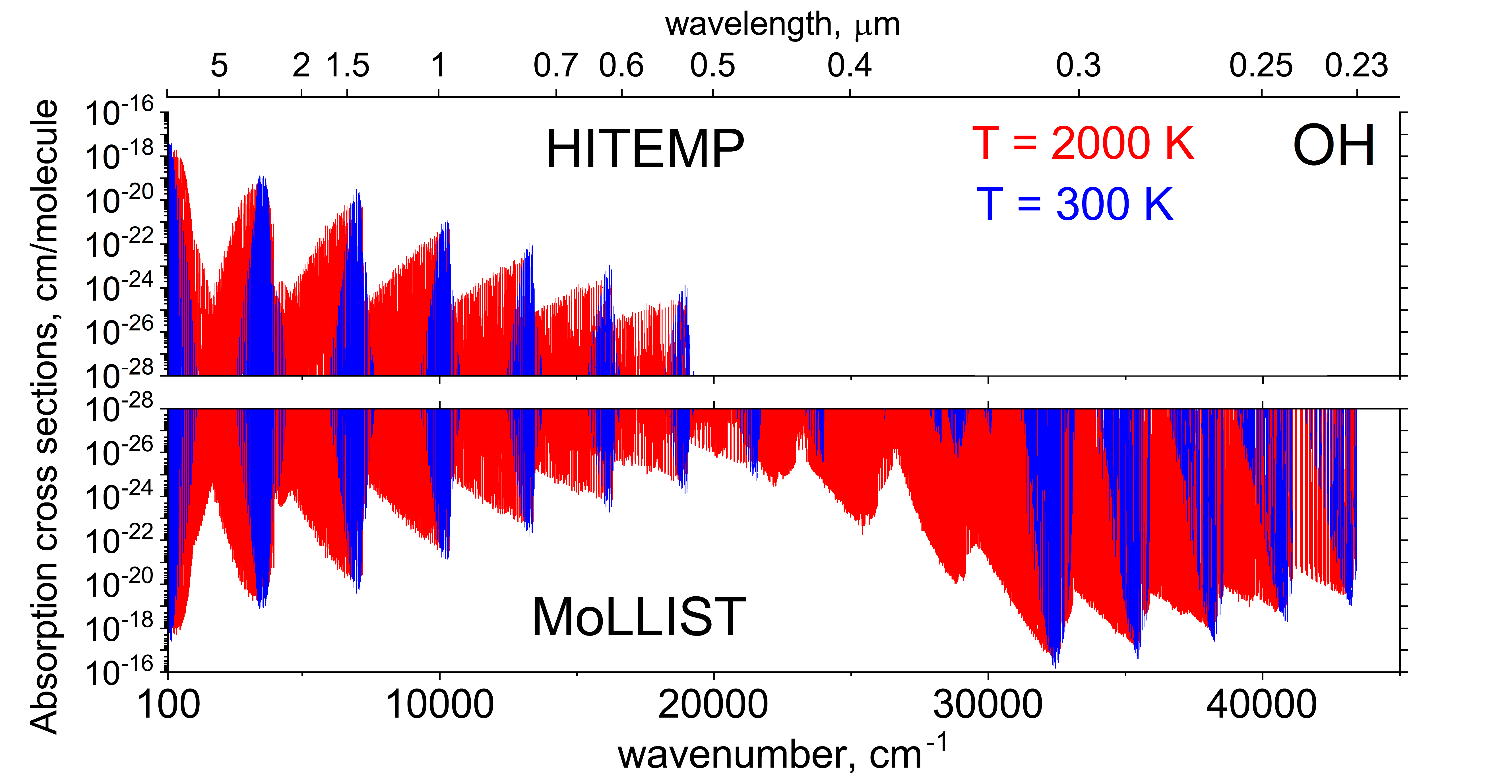}
\caption{Comparison of the HITEMP \citep{HITEMP2010} and MoLLIST line lists for OH: $T$ = 300 and 2000 K absorption cross sections computed using a Doppler line profile. The~MoLLIST spectrum is based on the data from \citet{16BrBeWe.OH}. }
\label{f:OH:MOL-HITEMP}
\end{figure}

The OH line list  was constructed from two empirical line lists. A~line list for the ground electronic state $X\,^2\Pi$ of OH was  produced by \citet{16BrBeWe.OH} for the vibrational (Meinel system) and pure rotational transitions covering $v$ up to 13 with  $J_{\rm max}$ ranging from 9.5 and 59.5, depending on the band. A~fit to the molecular constants was performed. This fit was based mainly on that
\mbox{of~\citet{09BeCoxx.OH}} but~included some new rotational data from \citet{11MaPiBa.OH}. The~absolute transition strengths were based on a new DMF, produced from a combination of two high level ab initio methods. A~line list for the $A\,^2\Sigma^+ - X\,^2\Pi$ electronic band system of OH was  calculated by \citet{18YoBeHo.OH}. Line positions were taken from \citet{94StBrAb.OH,80Coxon.OH,91CoSaCo.OH,05DePoDe.OH,92Yarkony.OH} and refitted with  \PGOPHER. Line intensities were calculated using a new ab initio transition dipole moment function (TDMF) obtained with Molpro 2012. This  new TDMF and PECs generated with program \RKR\ were used as input to  \Level\  which computed matrix elements of the transition dipole moment. The~quantum numbers are $J$, $v$, $F_1/F_2$ (spin component), and~the $e/f$ parity. The~line list is illustrated in Figure~\ref{f:OH:MOL-HITEMP}.

Both HITRAN \citep{jt691} and HITEMP \citep{HITEMP2010} provide line lists for OH;
the line lists only cover the ground electronic state $X$--$X$ transitions  and do not
use the work of \citet{16BrBeWe.OH}, as~also illustrated in Figure~\ref{f:OH:MOL-HITEMP}. In~particular, HITEMP was based on the older experimental data by \citet{98GoScGo.OH}. The~MoLLIST line list is both more accurate and more complete.

\subsection{OH$^+$}

\begin{figure}[h!]
\centering
\includegraphics[width=0.8\textwidth]{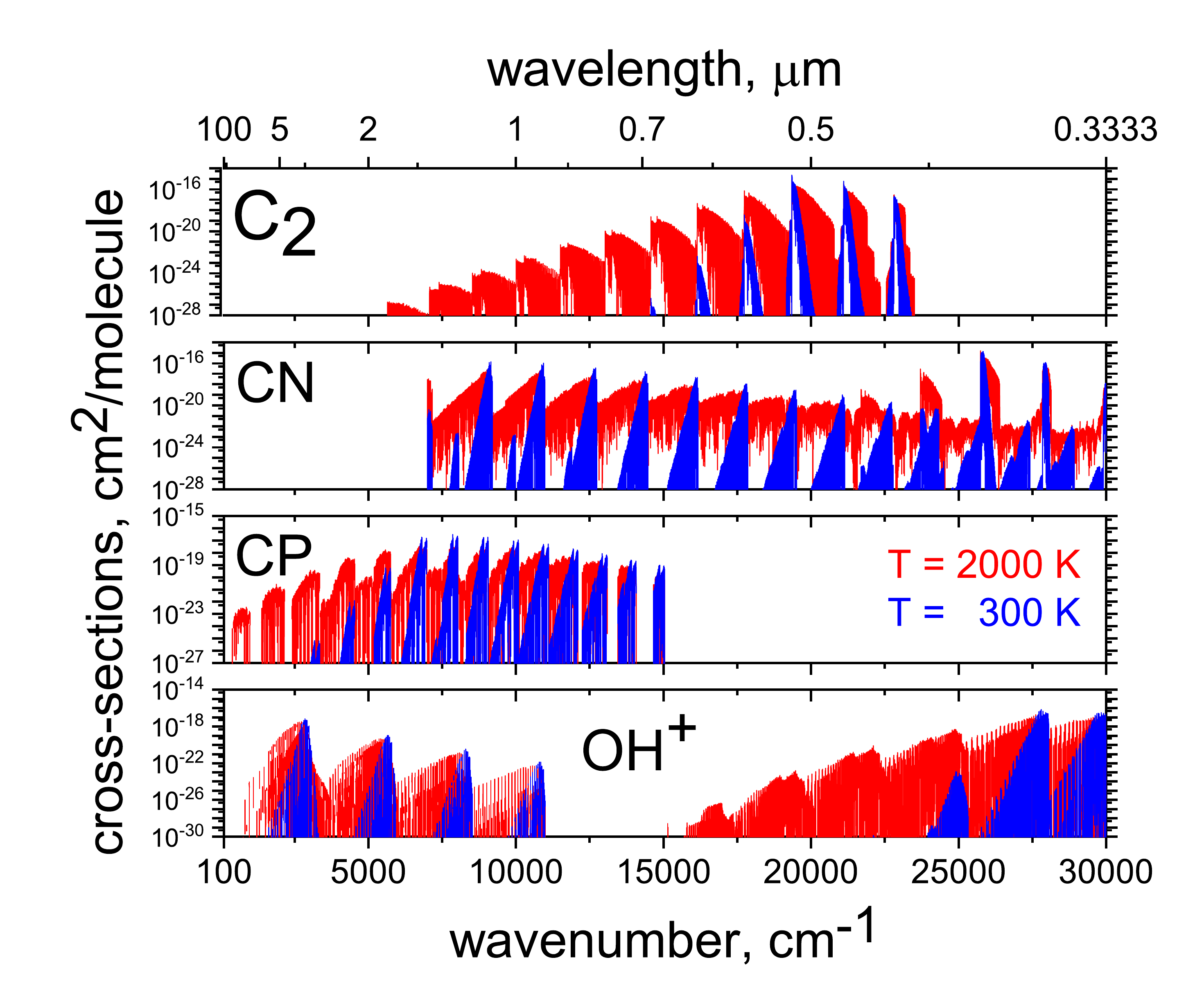}
\caption{Absorption cross sections of C$_2$ \citep{13BrBeSc.C2}, CN \citep{14BrRaWe.CN}, CP \citep{14RaBrWe.CP} and OH$^+$ \citep{17HoBexx.OH+,18HoBiBe.OH+}.}
\label{f:C2-CN-CP-OH}
\end{figure}

 An OH$^+$ line list for the $A\,^3\Pi-X\,^3\Sigma^-$ band system was produced by {\citet{17HoBexx.OH+,18HoBiBe.OH+}}. \citet{17HoBexx.OH+} fitted ground state rotational \citep{85BeVeMe.OH+,87LiHoOk.OH+} and rovibrational \citep{92ReJaXu.OH+,16MaHoPe.OH+} data sets using SPCAT/SPFIT \citep{91Pickett.OH+} with  Dunham constants from the Cologne Database for Molecular Spectroscopy (CDMS; \citet{16EnScSc.OH+}). The~ground state data were combined with the near-UV and predissociation data \citep{07RoBaSa.OH+} and fitted using \PGOPHER.  \citet{18HoBiBe.OH+} used their analysis  of a laboratory spectrum \citep{17HoBexx.OH+} with ab~initio methods to calculate infrared and ultraviolet oscillator strengths. These new oscillator strengths include branch dependent intensity corrections, arising from the Herman--Wallis effect. The~quantum numbers used are $J$, $v$, Electronic state, $F$ and $e/f$. The~line list is illustrated in Figure~\ref{f:C2-CN-CP-OH}.

\subsection{CaF}

\begin{figure}[h!]
\centering
\includegraphics[width=0.8\textwidth]{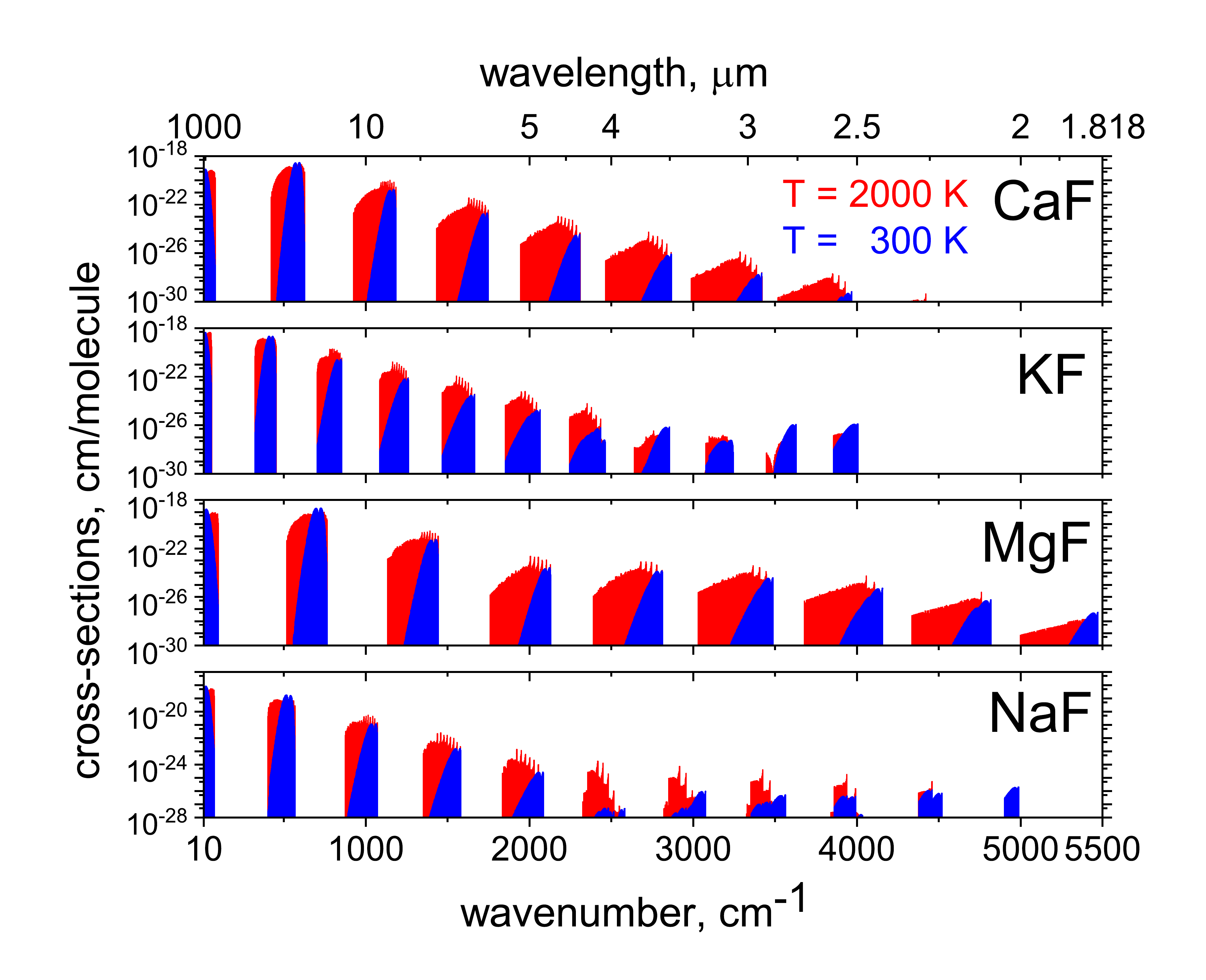}
\caption{Absorption cross sections of CaF \citep{18HoBexx.CaF}, KF and NaF \citep{16FrBeBr.NaF}, MgF \citep{17HoBexx.MgF}.}
\label{f:CaF:KF:MgF}
\end{figure}

A line list for the ground state of CaF  due \citet{18HoBexx.CaF} was based on  RCCSD(T) (restricted coupled clusters singles,
doubles and approximate triples) ab initio calculations. The~RCCSD(T) potential function was represented using the extended Morse oscillator (EMO) potential function and then refined by fitting to the observed lines positions of CaF from \citet{95ChGuZh.CaF}. With~the EMO potential and the RCCSD(T) dipole moment function, line lists for six isotopes of CaF were computed for $v \leq 10$, $J \leq 123$, $\Delta v = 0-10$. Finally, the~line lists were compiled into the ExoMol format. Spectra of CaF are illustrated  in Figure~\ref{f:CaF:KF:MgF}. The~quantum numbers used are $J$ and $v$.

\subsection{KF and~NaF}

Rotation--vibration line lists for NaF and KF in their ground electronic states were produced by \citet{16FrBeBr.NaF}. Experimental data used were previously measured for infrared transitions up to $v=8$ and $v=9$ for KF and NaF by \citet{96LiMuZh.KF} and \citet{96MuGuBe.NaF}. Additionally, for~NaF, three  microwave \citep{63BaLexx.NaF} and ten mm-wave \citep{65VeGoxx.NaF} lines were used to construct the global line list used for fitting.  Six $^{39}$KF microwave lines \citep{60GrLexx.KF,72DiFlGr.KF} and fourteen mm-wave lines \citep{65VeGoxx.NaF}  were used to construct the global line lists.  DMFs  were generated from ab initio calculations using the SA-CASSCF (state-averaged complete active space self-consistent field) and ACPF (averaged coupled-pairs functional) methods. PECs were determined by fitting various models to experimental data using the program \RKR\  and then \Level\  to generated ro-vibrational levels. \PGOPHER\ was used to calculate Einstein A values and line positions. The~quantum numbers used are $J$ and $v$. The~spectra of NaF and KF are illustrated in Figure~\ref{f:CaF:KF:MgF}.

\subsection{MgF}

Line lists for MgF in its $X$~$^2\Sigma^+$ ground states were generated by \citet{17HoBexx.MgF}. An~EMO PEC was obtained by fitting to observed laboratory vibration--rotation and pure rotational transitions from \citet{95BaZhGu.MgF}. Line lists were computed for $v \leq 8$, $J$, $\Delta v=0-8$ using the EMO PEC and an analytic DMF in the form of a Pad\'e approximant fitted to ab initio dipole moment data. Finally, the~line lists were compiled into ExoMol format (states and trans file), and~partition functions were calculated according to \citet{84SaTaxx.partfunc}. The~quantum numbers used include $J$ and $v$. The~spectrum of MgF is illustrated in Figure~\ref{f:CaF:KF:MgF}.

\subsection{LiCl and~LiF}

\begin{figure}[h!]
\centering
\includegraphics[width=0.8\textwidth]{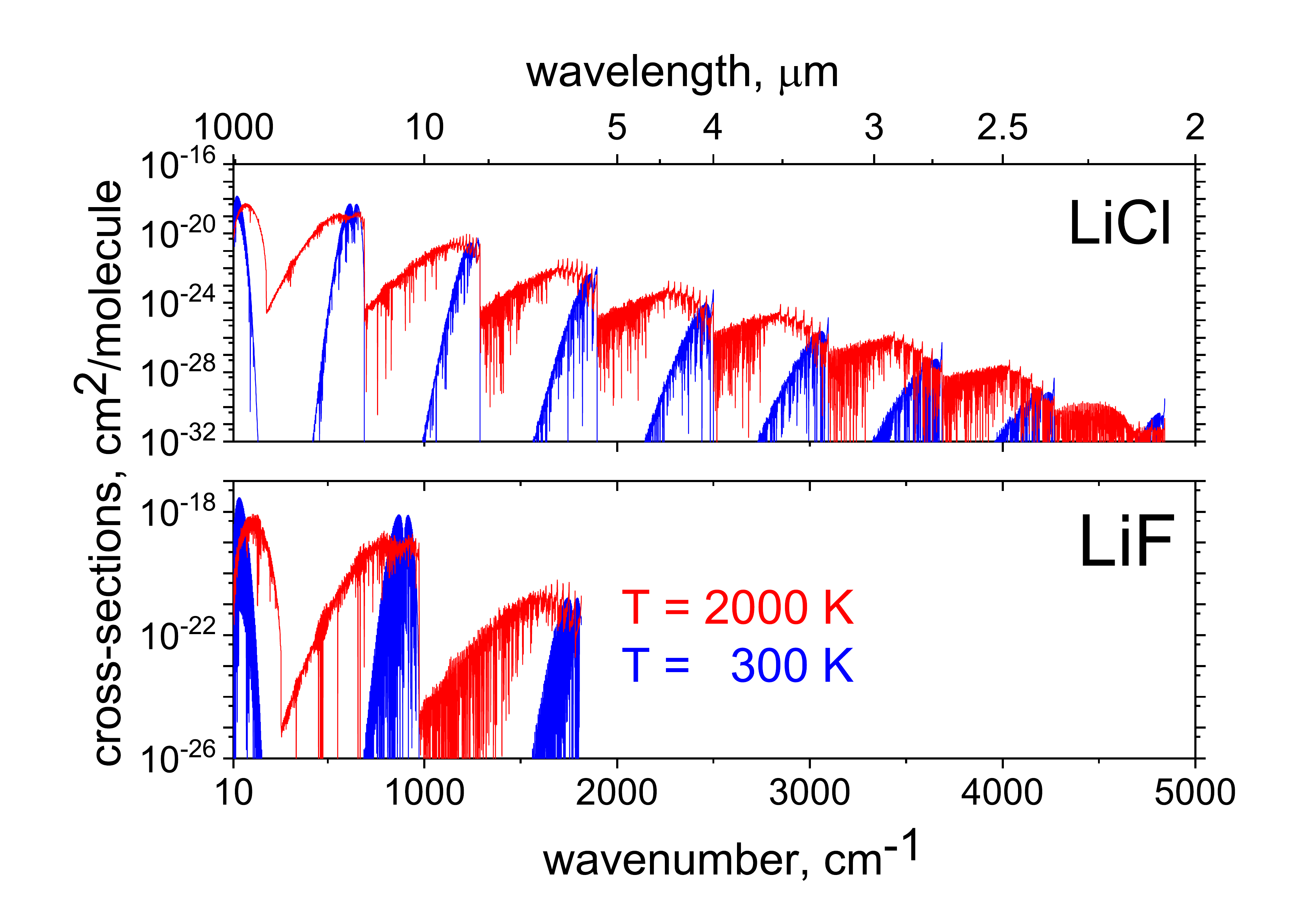}
\caption{Absorption cross sections of LiCl and LiF from \citet{18BiBexx.LiF}.}
\label{f:LiCl-LiF}
\end{figure}

Ro-vibrational line lists for LiF and LiCl in their ground $X\,^1\Sigma^+$  electronic states were computed  by \citet{18BiBexx.LiF}. The~ro-vibrational energy levels were calculated using empirical EMO PECs which were determined by direct potential-fitting to the new data from \citet{18BiBexx.LiF}
using dPotFit \citep{dPotFit}.
Ab initio DMCs were obtained using  Molpro 2012 with the MRCI/aug-cc-pwCV5Z level of theory and used for the line strength calculations.  Program \Level\  was employed to calculate the
final line lists. The partition functions of LiF and LiCl were taken from \citet{18BiBexx.LiF} and multiplied by the corresponding statistical nuclear weighs, according to the physicists convention.
The~quantum numbers used are $J$ and $v$. Spectra computed using  these line lists are illustrated in Figure~\ref{f:LiCl-LiF}.

\subsection{MgH}

Ro-vibrationally resolved transitions for the $A\,^2\Pi \rightarrow X\,^2\Sigma^+$ and $B'\,^2\Sigma^+ \rightarrow X\,^2\Sigma^+$ band systems of $^{24}$MgH were calculated by \citet{13GhShBe.MgH}. Empirically-determined analytic PECs \citep{11ShBexx.MgH} for the $X\,^2\Sigma^+$ state and RKR potentials for the $A\,^2\Pi$ and $B'\,^2\Sigma^+$ states  were combined with the most recent ab initio TDMCs computed by \citet{12MoShxx.MgH} to produce a rovibronic line list. The program \Level\ was used to calculate transition dipole moment and \PGOPHER\  was used to generate the line list. The~calculated rovibronic transition frequencies were improved by substituting the term values of the $X$~$^2\Sigma^+$, $A\,^2\Pi$ and $B'\,^2\Sigma^+$ states of $^{24}$MgH with the empirical values from \citet{07ShHeLe.MgH}. Using the ab initio TDMCs, transition frequencies and H\"{o}nl--London factors, the~Einstein~$A$ coefficients were calculated individually for $\sim$30,000 rovibronic lines of the $A\,^2\Pi \rightarrow X\,^2\Sigma^+$ and $B'\,^2\Sigma^+ \rightarrow X\,^2\Sigma^+$ systems of $^{24}$MgH. The~quantum numbers used are $J$, electronic state, $v$, $N$ (total angular momentum quantum number excluding spin), $F_1/F_2$, and $e/f$.
The spectra computed using  these line lists are illustrated in  Figure~\ref{f:MgH:TiH}.

Note that there are existing ExoMol line lists for isotopologues of MgH which cover rotation--vibration transitions within the ground electronic state \citep{jt529}.

\subsection{TiH}

\begin{figure}[h!]
\centering
\includegraphics[width=0.8\textwidth]{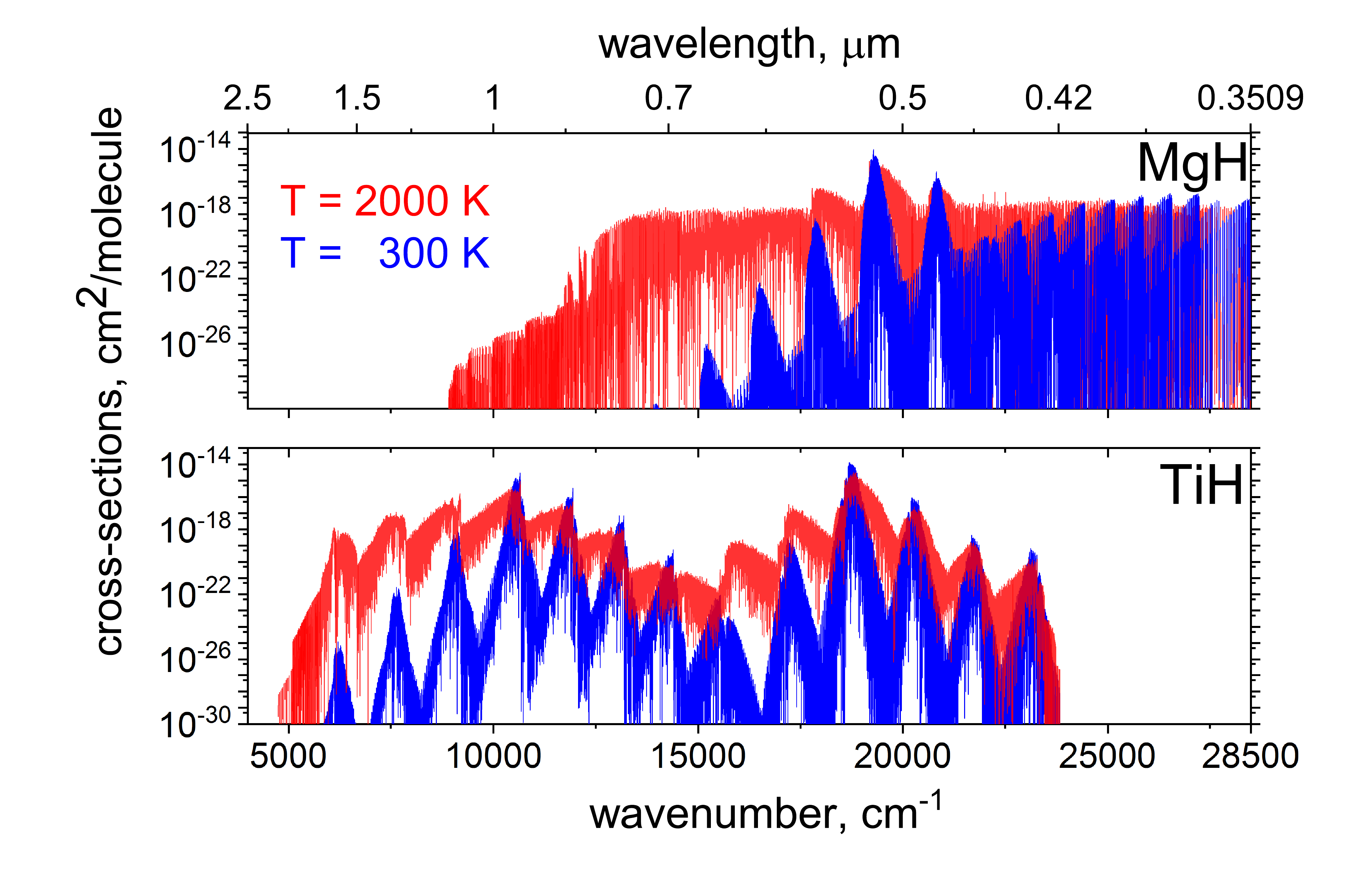}
\caption{Absorption cross sections of MgH and TiH from \citet{13GhShBe.MgH} and \citet{05BuDuBa.TiH}.}
\label{f:MgH:TiH}
\end{figure}

A line list for the  TiH $B\,^4\Gamma - X\,^4\Phi$ and $A\,^4\Phi - X\,^4\Phi$ systems were produced by \citet{05BuDuBa.TiH} using a combination of the experimental line positions \citep{91StShSi.TiH,96LaLixx.TiH,03AnBaBe.TiH,03AnBaLi.TiH} and ab initio transition probabilities.  The~latter were obtained based on Molpro 2002 calculations  using the IC-MRCI/SA-CASSCF level of theory; scalar relativistic effects were accounted for using the Douglas--Kroll--Hess (DKH) approach. The~quantum numbers used are $J$, $v$, $\Omega$, $+/-$ parity, electronic state. Illustrative spectra of TiH computed using  this line list are given in  Figure~\ref{f:MgH:TiH}.

\subsection{CrH}

\begin{figure}[h!]
\centering
\includegraphics[width=0.8\textwidth]{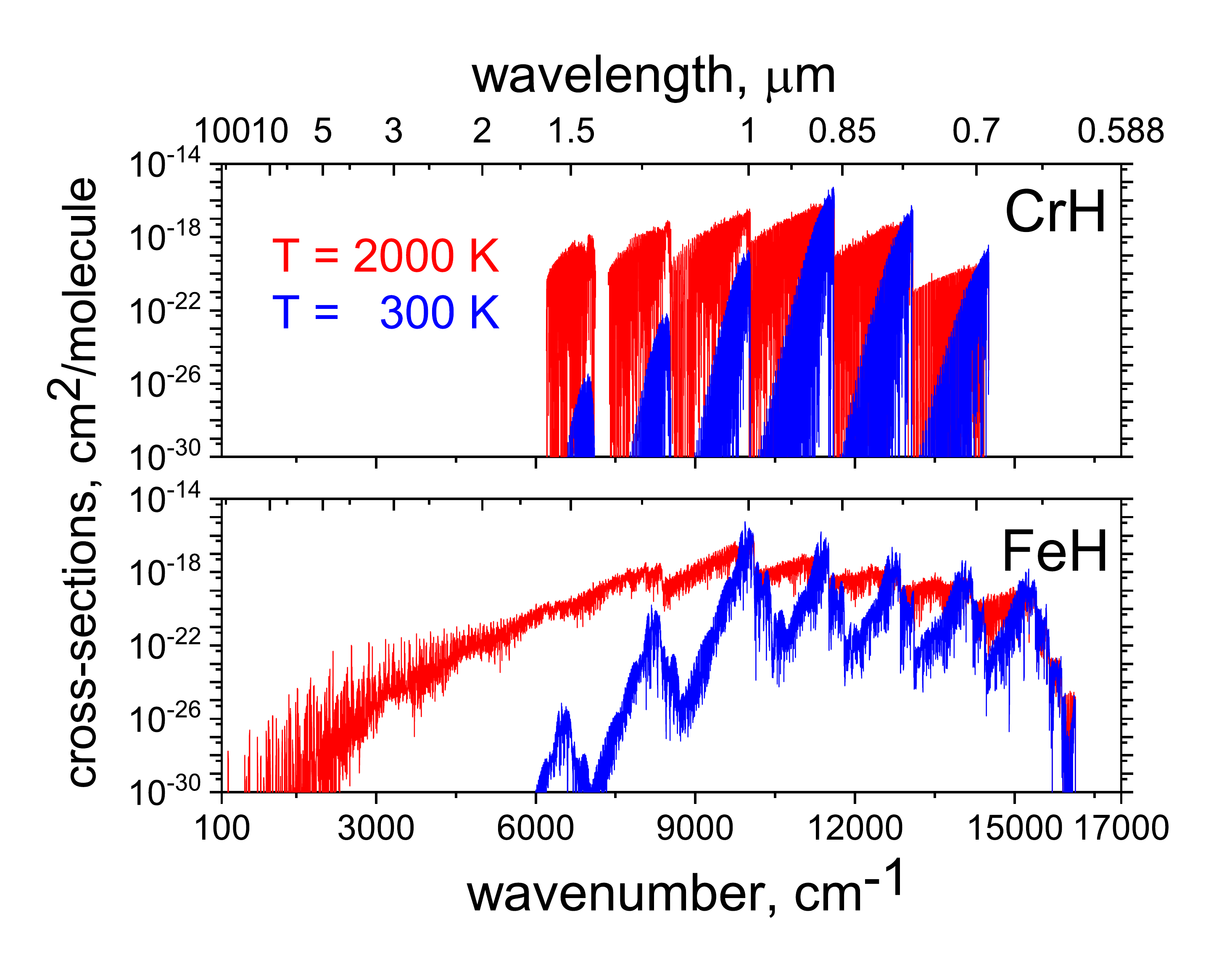}
\caption{Absorption cross sections of CrH from \citet{06ChMeRi.CrH} and FeH from \citet{03DuBaBu.FeH}.}
\label{f:CrH:FeH}
\end{figure}

A limited line list for CrH was provided by \citet{06ChMeRi.CrH}, see Figure~\ref{f:CrH:FeH}. The~(1,0) band of the $A\,^6\Sigma^+ - X\,^6\Sigma^+$ system  was observed and rotational assignments for levels with $N \leq 3$  made. These assignments were based on
Fourier transform emission spectra~\cite{01BaRaBe.CrH} for which higher-$N$ lines were assigned. The~low-$N$ rotational levels are extensively perturbed, presumably by levels of the $a\,^4\Sigma^+$, $v=1$ and $B\,^6\Pi$, $v=0$ states. The~quantum numbers used are $J$, $v$, $N$, $e/f$ parity, electronic~state.

\subsection{FeH}

A  $F\,^4\Delta - X\,^4\Delta$  line list ($v=0,1,\ldots,4$, $J\le 50.5$) for FeH was constructed by \citet{03DuBaBu.FeH} based on the experimental energy term values from \citet{87PhDaLi.FeH} ($v=0,1,2$) and an ab~initio (IC-MRCI+Q) DTMF. The~rovibronic energies were obtained using empirical spectroscopic constants $T_v$, $B_v$, $A_v$ and $\lambda_v$. The~$F$--$X$ Einstein coefficients were computed using H\"onl--London factors and vibrational line strengths based on RKR potentials.  The~line list contains 25 vibronic bands $v',v" = 0,1,2,3,4$ in the $F\,^4\Delta - X\,^4\Delta$ system of FeH. The~quantum numbers used are $J$, $v$, $\Omega$, $e/f$ and electronic states labels (`X4Delta' and `F4Delta'). The~FeH spectra simulated using the line list by \citet{03DuBaBu.FeH} is illustrated in Figure~\ref{f:CrH:FeH}.




\subsection{C$_2$}

\citet{13BrBeSc.C2} calculated rotational line strengths for the C$_2$ Swan system ($d\,^3\Pi_g - a\,^3\Pi_u$)  for bands with $v'=0-10$ and $v''=0-9$ with $J_{\rm max}$ ranging from 34 to 96, depending on the band using previous observations in 33 vibrational bands \citep{85CuSaxx.C2,85SuSaHi.C2,94PrBexx.C2,02TaAmxx.C2,07TaHiAm.C2}.  Line strengths were based on an ab initio calculation of DTMC using MRCI/aug-cc-pV6Z, taking into account core and core-valence (CV) correlation corrections computed using the aug-cc-pCVQZ basis set and the scalar relativistic energy corrections evaluated via the DKH approach in conjunction with the appropriate cc-pVQZ basis sets. Transition moments were computed by using a bi-orthogonal transformation of the mutually non-orthogonal orbitals of the two states. The~quantum chemical calculations were performed using Molpro 2006. The~potential energy curves were calculated using the computer program \RKR. \PGOPHER\ was used to calculate Einstein $A$ values and line positions. Rotationless TDMs were calculated using the computer program \Level.  The~quantum numbers used are $J$, $F$, $e/f$, $v$, and electronic state. This line lists were converted to ExoMol format by \citet{18ViSmPr}. The~C$_2$ spectra of \citet{13BrBeSc.C2} simulated using this line list is illustrated in Figure~\ref{f:C2-CN-CP-OH}.

The ExoMol database already contains a line list for C$_2$:  \citet{jt736}
constructed a line list, called 8states, because it consists of transitions between the lowest eight electronic
states, as~part of the ExoMol project. It was produced using a combination of an empirical spectroscopic model and high level ab initio TDMCs. Nuclear
motion calculations were performed with the program \Duo\  \citep{jt609}
which allows for full treatment of spin-orbit and other couplings. The~8states line list was improved by replacing the theoretical energies with experimentally determined MARVEL energy values of C$_2$~\citep{jt637} and is thus partly based on the same experimental data as the Swan line list of \citet{13BrBeSc.C2}. 8states is more complete than that of \citet{13BrBeSc.C2} in
that it considers all transitions with $J \leq 190$ for not only the Swan system, but
also seven other band systems. For~most purposes, 8states provides a more appropriate starting point
and the MoLLIST of \citet{13BrBeSc.C2} is included in case it is needed for specialist
applications.

\subsection{CP}

\citet{14RaBrWe.CP} calculated line strengths for transitions in the $A\,^2\Pi - X\,^2\Sigma^+$ band system of CP, which include the effect of rotation on the vibrational wavefunctions (the Herman--Wallis effect), using the programs \PGOPHER\ and \Level, based on experimental data due to \citet{87RaBexx.CP,92RaTaBe.CP,89SaYaKa.CP,99KlKlWi.CP}. RKR PECs for the $A$~$^2\Pi$ and $X$~$^2\Sigma^+$ states were  constructed using spectroscopic constants  from high resolution spectra. The~RKR potentials of the two states, and~ab initio electronic TDMFs of this from~\cite{98deFexx.CP} were used in \Level\ to produce transition dipole moment matrix elements. The~matrix elements where then converted from Hund's case (b) to (a), and used in \PGOPHER\ to generate a line list containing observed and calculated wavenumbers, Einstein $A$ coefficients and $f$-values for 75 bands with $v$=0--8 for both states.  Quantum numbers used are $J$, electronic state, $v$, $F$ and $e/f$.
Spectra of CP of simulated using this line list are illustrated in Figure~\ref{f:C2-CN-CP-OH}.

\subsection{CN}

The line list is due to \citet{14BrRaWe.CN}. RKR PECs for the $A$~$^2\Pi$, $B$~$^2\Sigma^+$, and~$X$~$^2\Sigma^+$ states were computed using spectroscopic constants from the $A\,^2\Pi - X\,^2\Sigma^+$ and $B\,^2\Sigma^+ - X\,^2\Sigma^+$ transitions based on the experimental data from \citet{10RaWaBe.CN,06RaDaWa.CN,91DaAbRa.CN,04HoCiSp.CN,14BrRaWe.CN}.
New electronic TDMFs for these systems and a dipole moment function for the $X$~$^2\Sigma^+$ state were generated from high level ab initio calculations and used in \Level\ to produce transition dipole moment matrix elements for a large number of vibrational bands. \PGOPHER\ was then used to calculate Einstein $A$ coefficients, and~a line list was generated containing the observed and calculated wavenumbers, Einstein $A$ coefficients and $f$-values for 290 bands of the $A^2\Pi - X^2\Sigma^+$ system with $v'$ = 0--22, $v''$ = 0--15, 250 bands of the $B^2\Sigma^+ - X^2\Sigma^+$ system with $v' = 0-15$, $v'' = 0-15$ and 120 bands of the ro-vibrational transitions within the $X$~$^2\Sigma^+$ state with $v$ = 0--15. The~quantum numbers used are $J$, electronic state, $v$, $F$ and $e/f$.
The CM spectra of \citet{14BrRaWe.CN} simulated using this line list is illustrated in Figure~\ref{f:C2-CN-CP-OH}.

\subsection{CaH}

\begin{figure}[h!]
\centering
\includegraphics[width=0.8\textwidth]{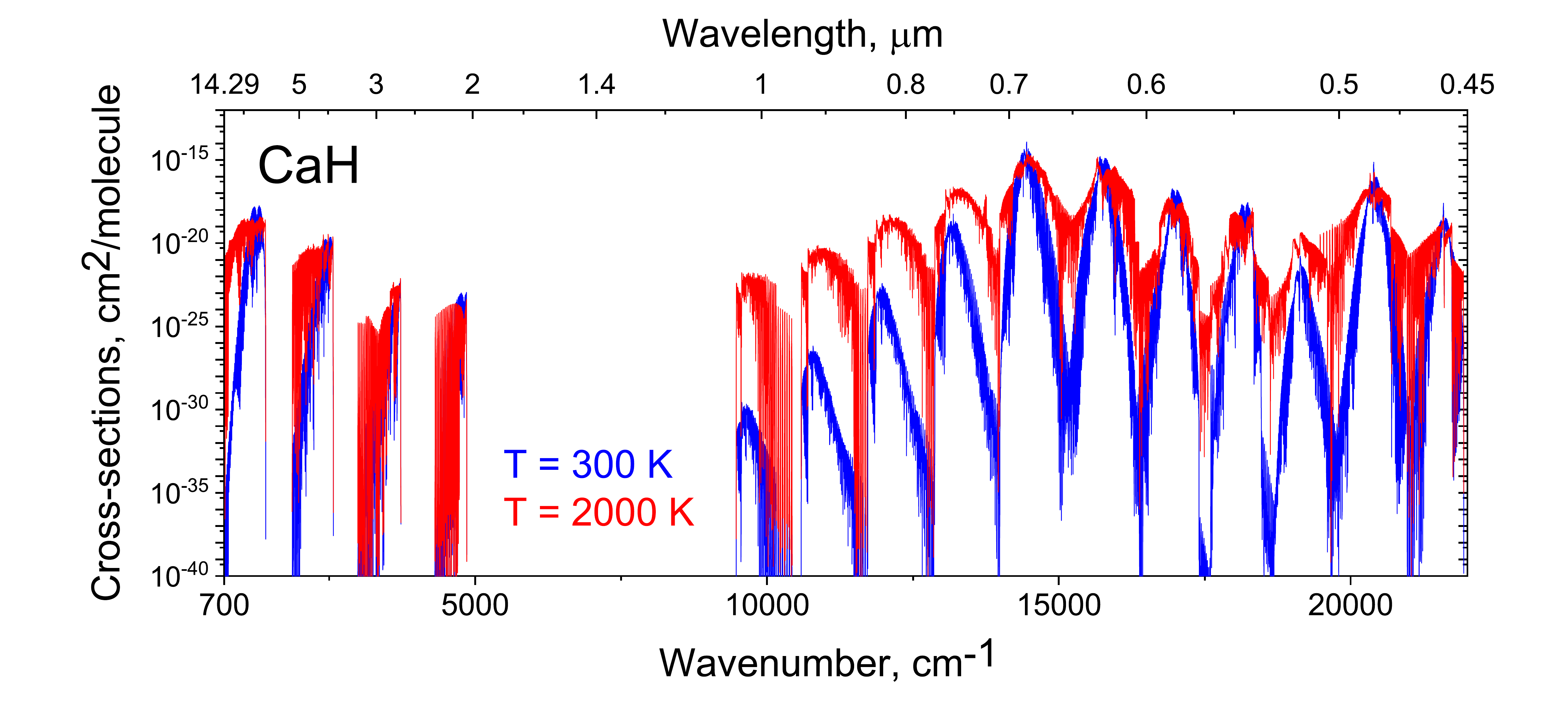}
\caption{Absorption cross sections of CaH from \citet{12LiHaRa.CaH,13ShRaBe.CaH,17FaShxx.CaH}.}
\label{f:CaH}
\end{figure}

The CaH line list ($X$--$X$, $A$--$X$, $B$--$X$, $E$--$X$) is based on the experiment of \citet{11RaTeGo.CaH,13ShRaBe.CaH}; see Figure~\ref{f:CaH}.
Einstein $A$ coefficients and absolute line intensities of $E\,^2\Pi - X\,^2\Sigma^+$ transitions of CaH were calculated by \citet{12LiHaRa.CaH}. Potential energy functions for both $E$~$^2\Pi$ and $X$~$^2\Sigma^+$ states of CaH were constructed using the purely numerical RKR method with the Kaiser correction and the effective Dunham coefficients from \citet{11RaTeGo.CaH}. These RKR potential curves and a high-level ab initio TDMF from \citet{03WeStKi.CaH}  were then employed in \Level\ to calculate a transition dipole moment matrix elements for the 10 bands involving $v'=0,1$ of the $E\,^2\Pi$ state and $v''= 0, 1, 2, 3, 4$ of the $X\,^2\Sigma^+$ state using the rotational line strength factors (H\"{o}nl--London factors)  derived for the intermediate coupling case between Hund's case (a) and (b). The~computed transition dipole moments and the spectroscopic constants from \citet{11RaTeGo.CaH} were combined to generate line lists containing Einstein $A$ coefficients and absolute line intensities for the $X\,^2\Sigma^+$--$X\,^2\Sigma^+$ and $E\,^2\Pi - X\,^2\Sigma^+$ systems of CaH for $J$-values up to 50.5. We  used the line positions from \citet{11RaTeGo.CaH,13ShRaBe.CaH} to reconstruct the energy levels using the MARVEL procedure \citep{jt412}. The~Einstein coefficients of the $A\,^2\Pi$--$X\,^2\Sigma^+$, $B\,^2\Sigma^+$--$X\,^2\Sigma^+$ system were taken from \citet{17FaShxx.CaH}.  The~quantum numbers used are $J$, $e/f$, $v$ and electronic~state.

Note that there is an existing ExoMol line list for  $^{40}$CaH which covers rotation--vibration transitions within the ground electronic state \citep{jt529}.

\subsection{N$_2$}

\begin{figure}[h!]
\centering
\includegraphics[width=0.8\textwidth]{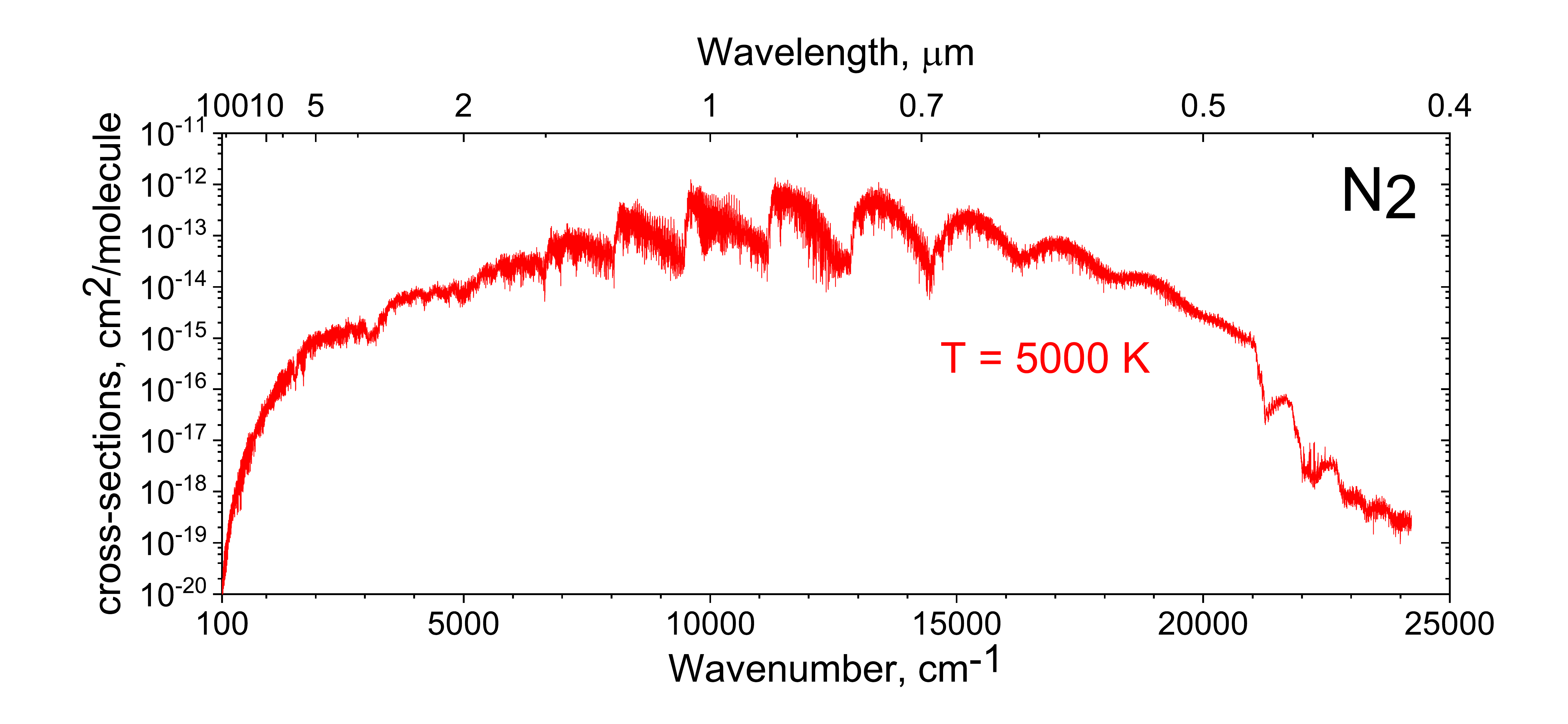}
\caption{Line lists for $N_2$ from \citet{18WeCaCr.N2}.}
\label{f:N2_5000K}
\end{figure}

The new line list for triplet states of N$_2$  was produced by \citet{18WeCaCr.N2} using experimental data from \citet{15BeRexx.N2}, see Figure~\ref{f:N2_5000K}; the line list includes the $B$~$^3\Pi_g$--$A$~$^3\Sigma_u^+$, $B'$~$^3\Sigma_u^-$--$B$~$^3\Pi_g$,  $W$~$^3\Delta_u$--$B$~$^3\Pi_g$ systems. Their analysis used high-level ab initio calculations of PECs, TDMFs, and spin-orbit coupling constants to prepare the model and extend the potential range of applicability.  Final line list calculations
were performed with \PGOPHER.
The partition function is based on energies of the singlet ($X$~$^1\Sigma_g^+$) ground electronic  state of N$_2$ computed using the temperature points of \citet{16BaCoxx.partfunc} interpolated on a 1~K grid from 0 to 10,000~K using cubic splines. The~ExoMol name of the line list is~WCCRMT.

\subsection{Lifetimes}

\begin{figure}[h!]
\includegraphics[width=0.33\textwidth]{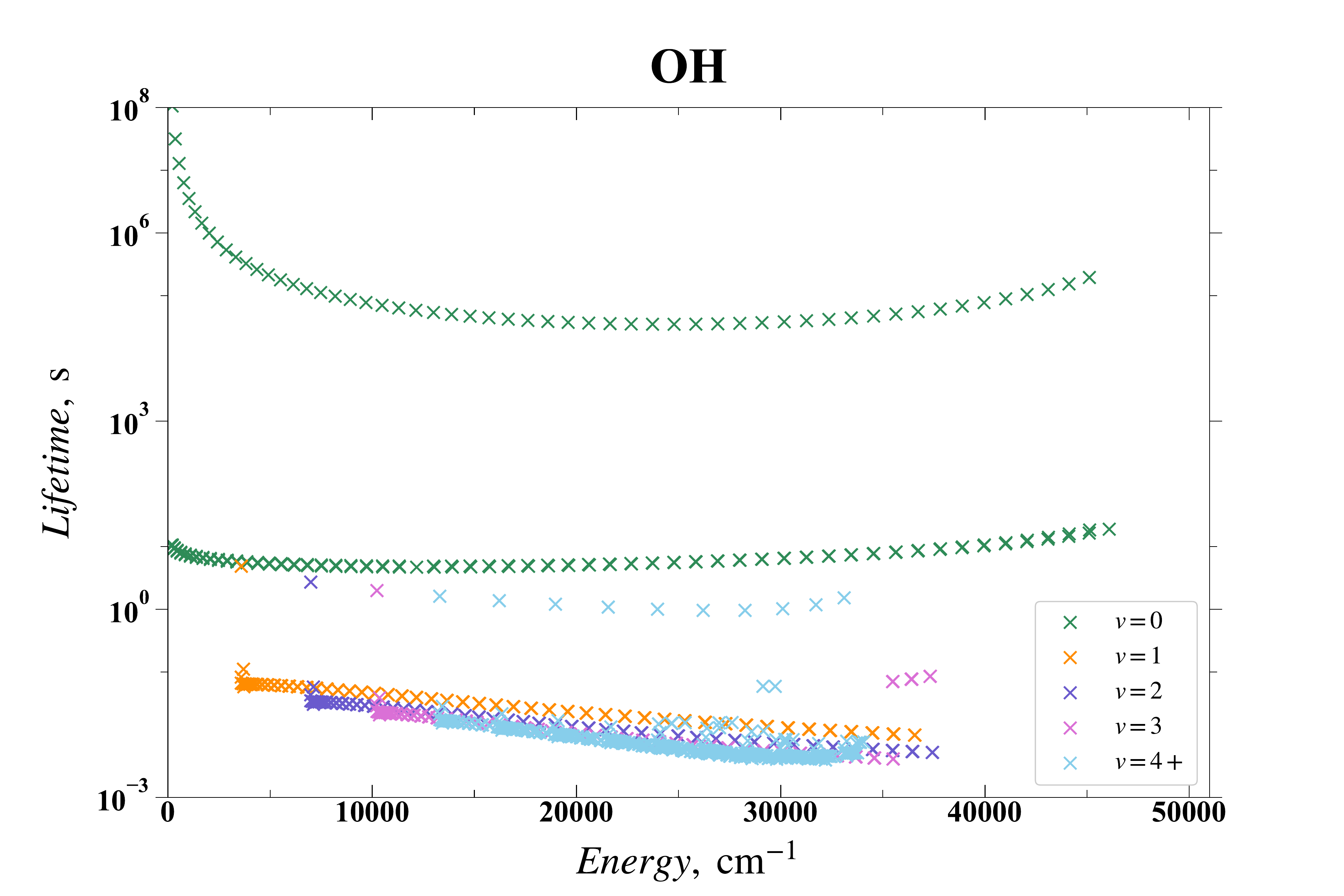}
\includegraphics[width=0.33\textwidth]{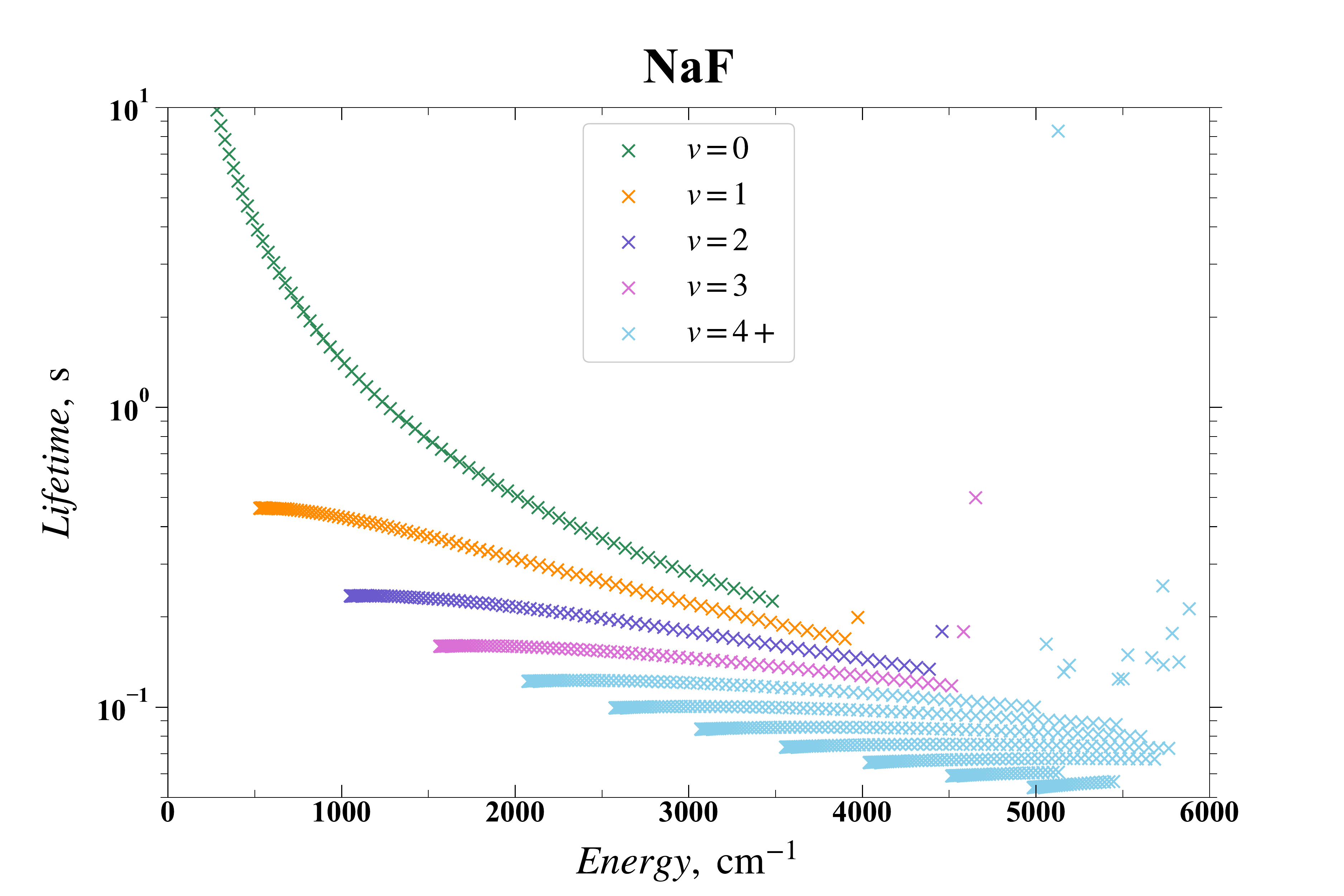}
\includegraphics[width=0.33\textwidth]{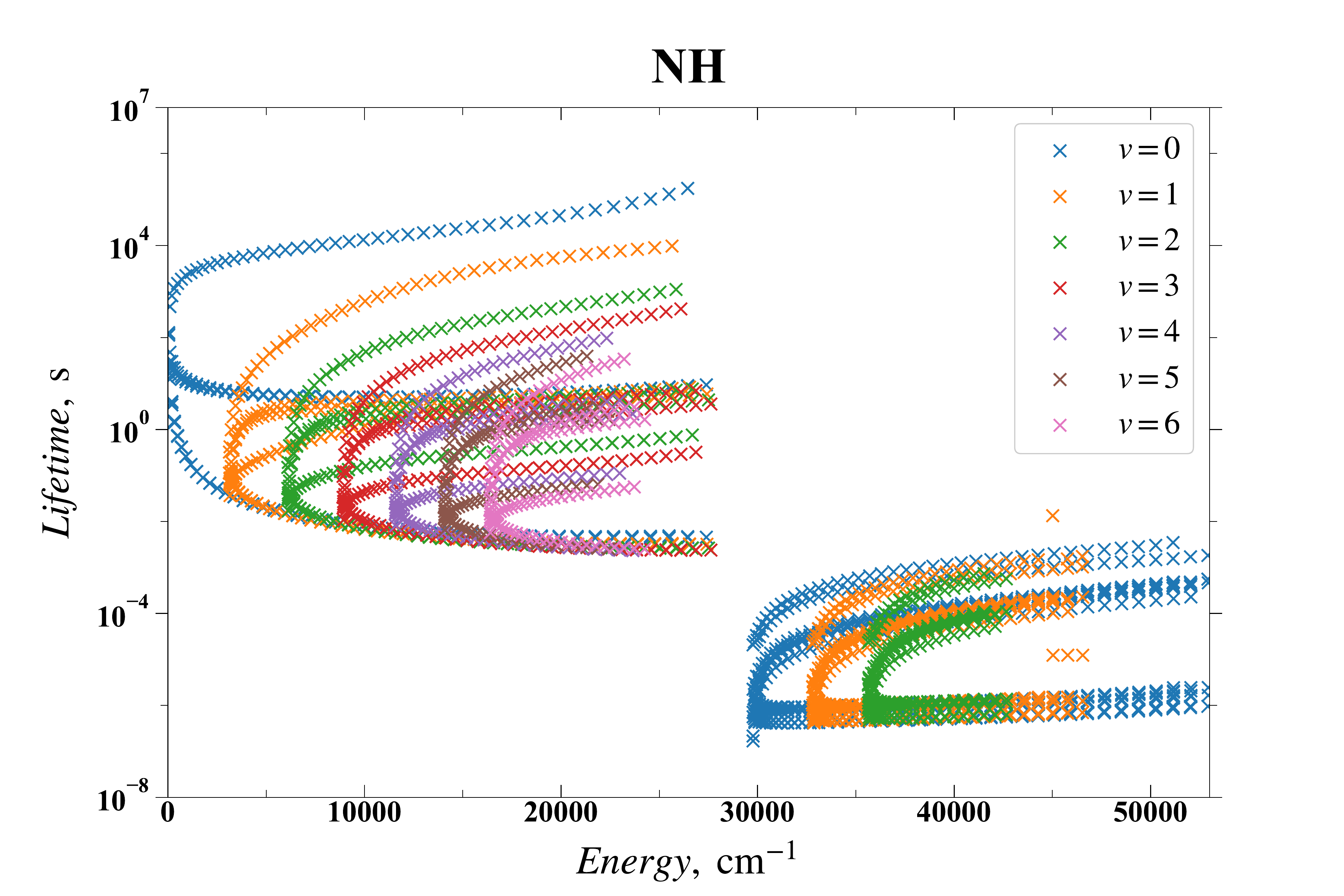}
\includegraphics[width=0.33\textwidth]{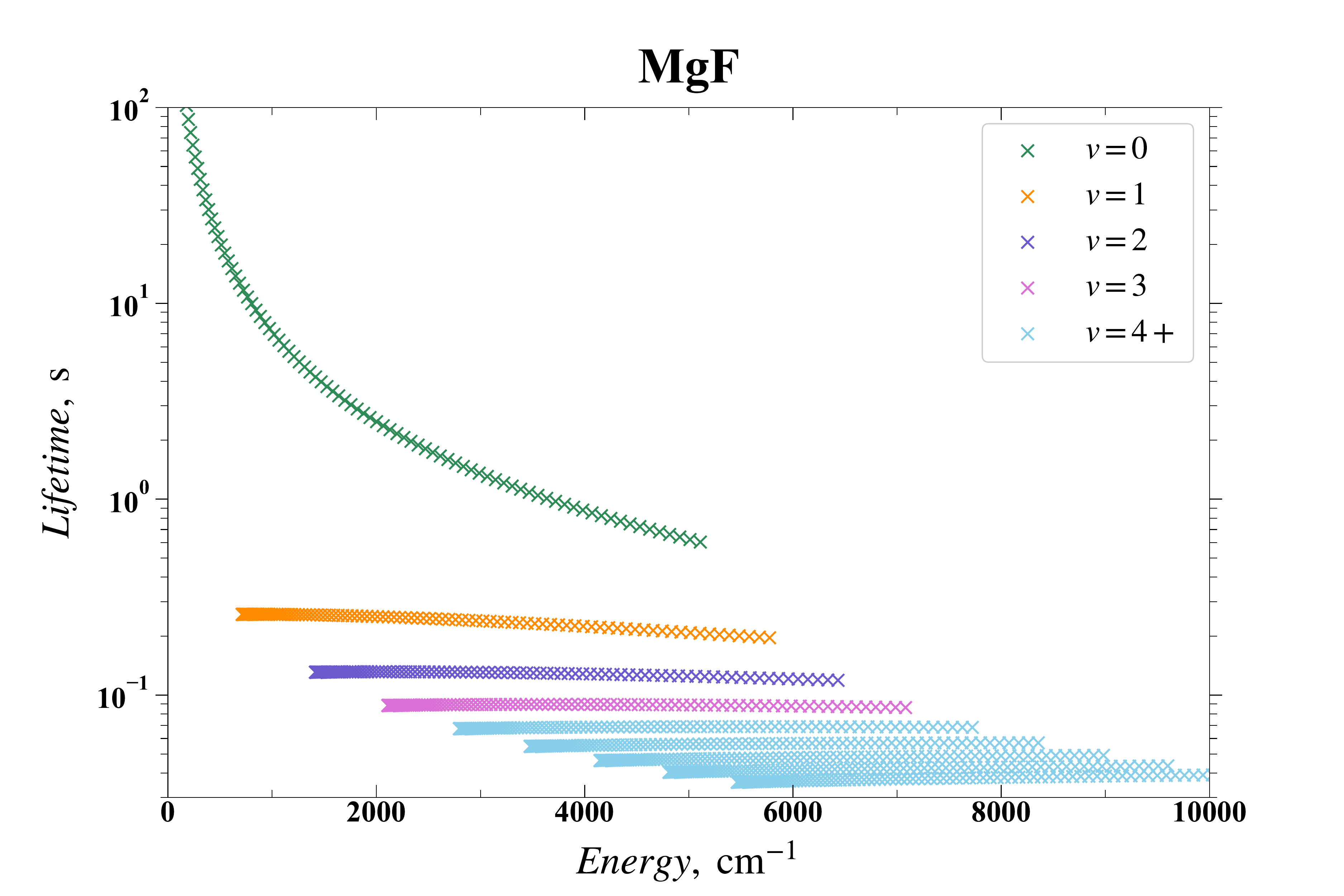}
\includegraphics[width=0.33\textwidth]{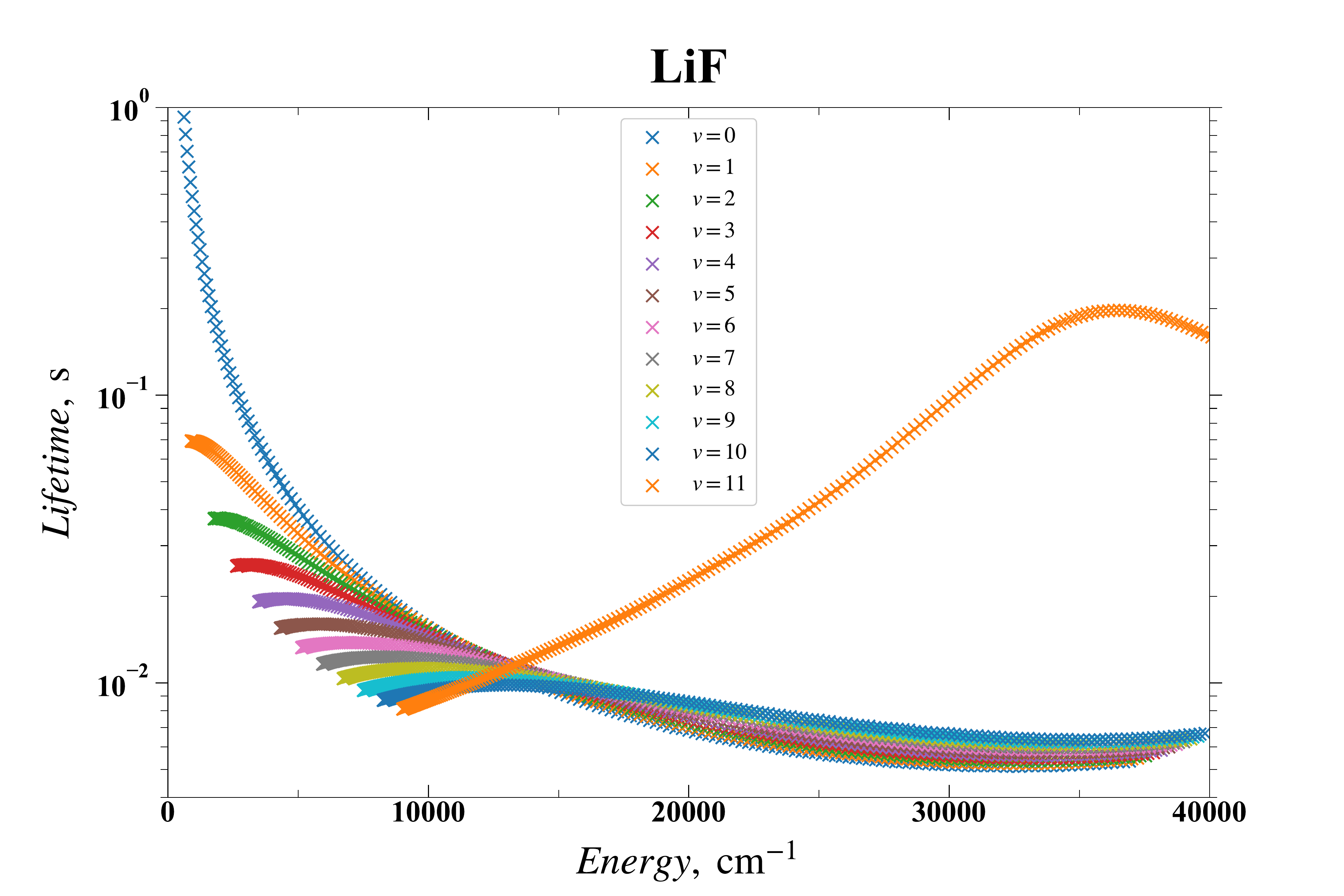}
\includegraphics[width=0.33\textwidth]{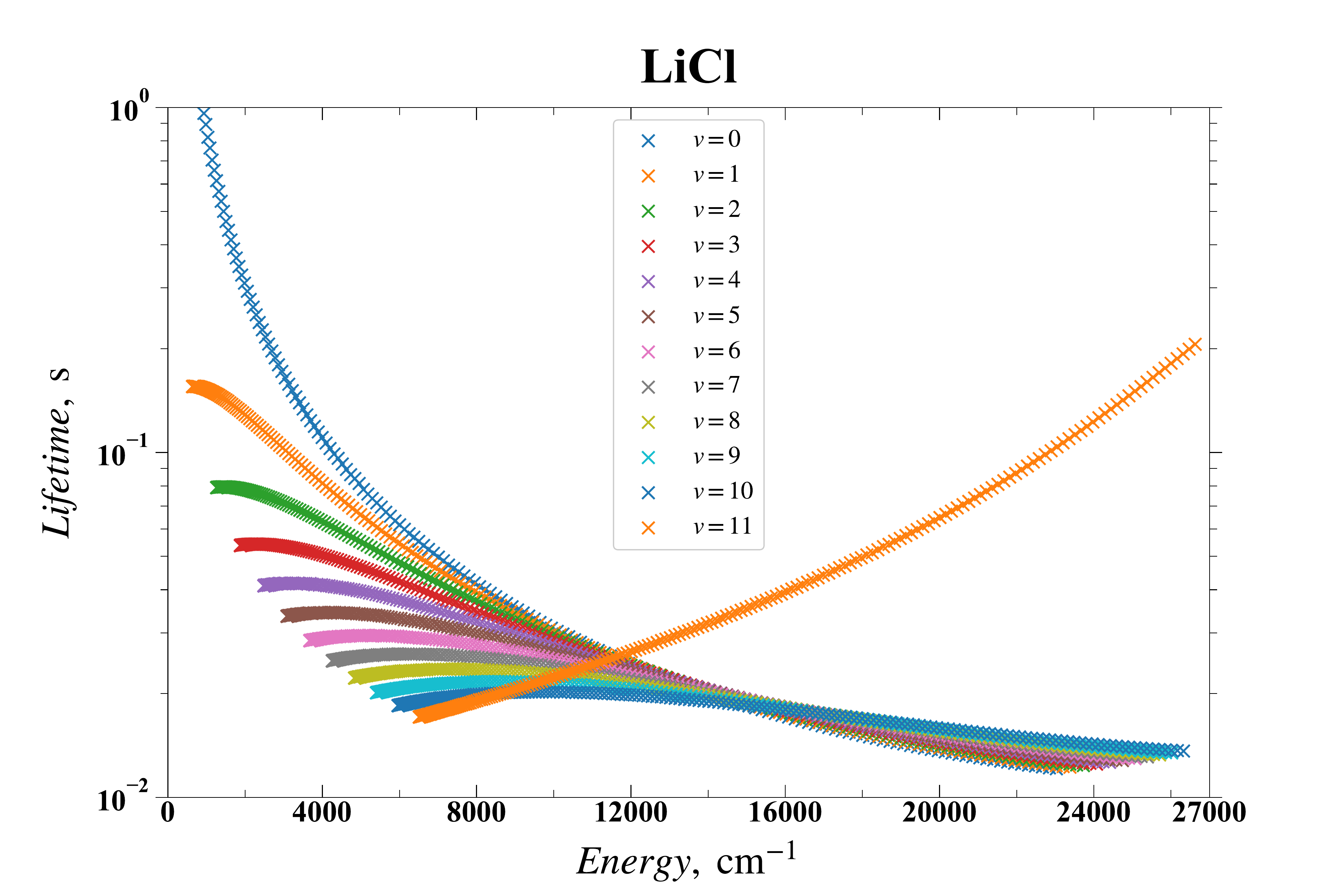}
\includegraphics[width=0.33\textwidth]{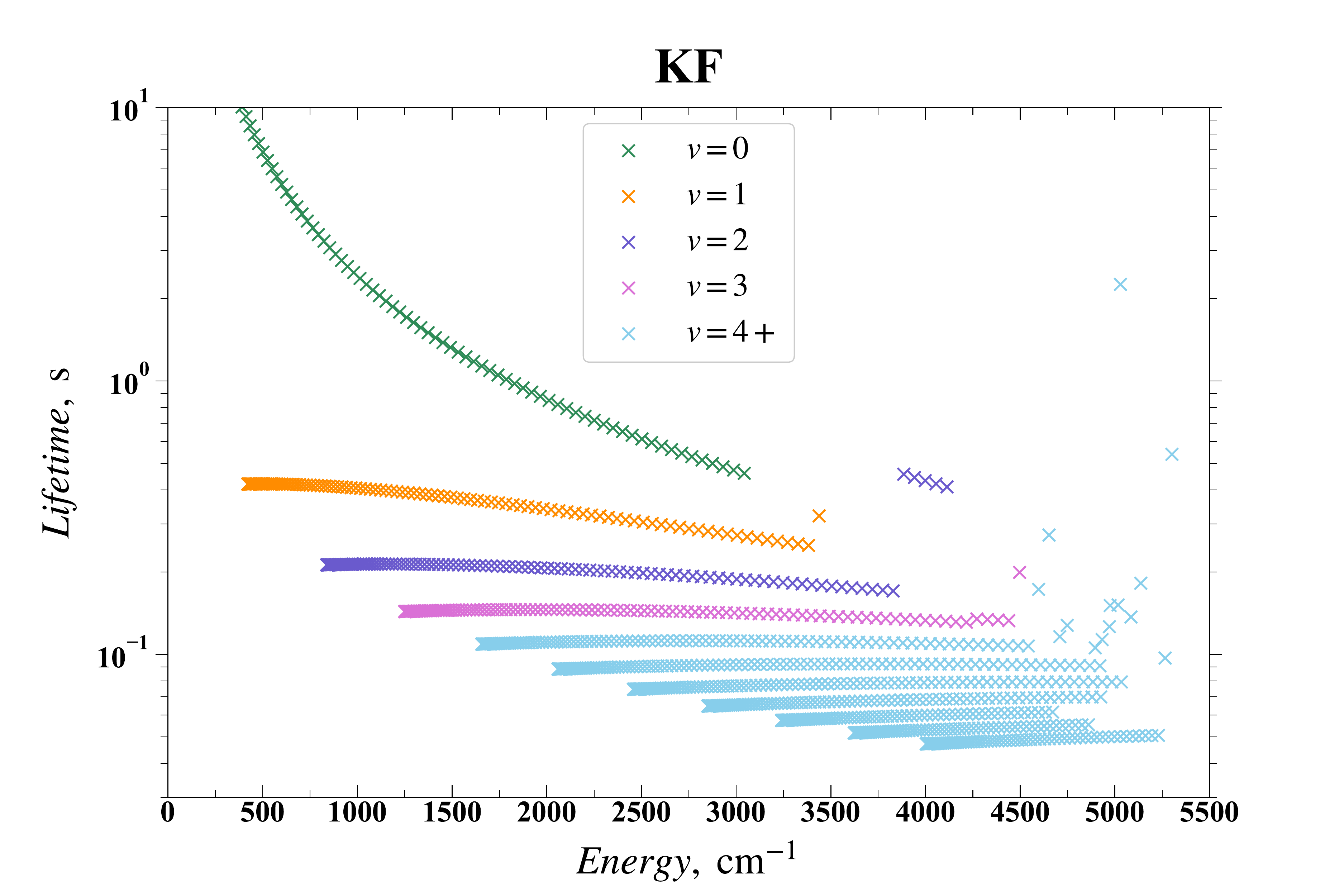}
\includegraphics[width=0.33\textwidth]{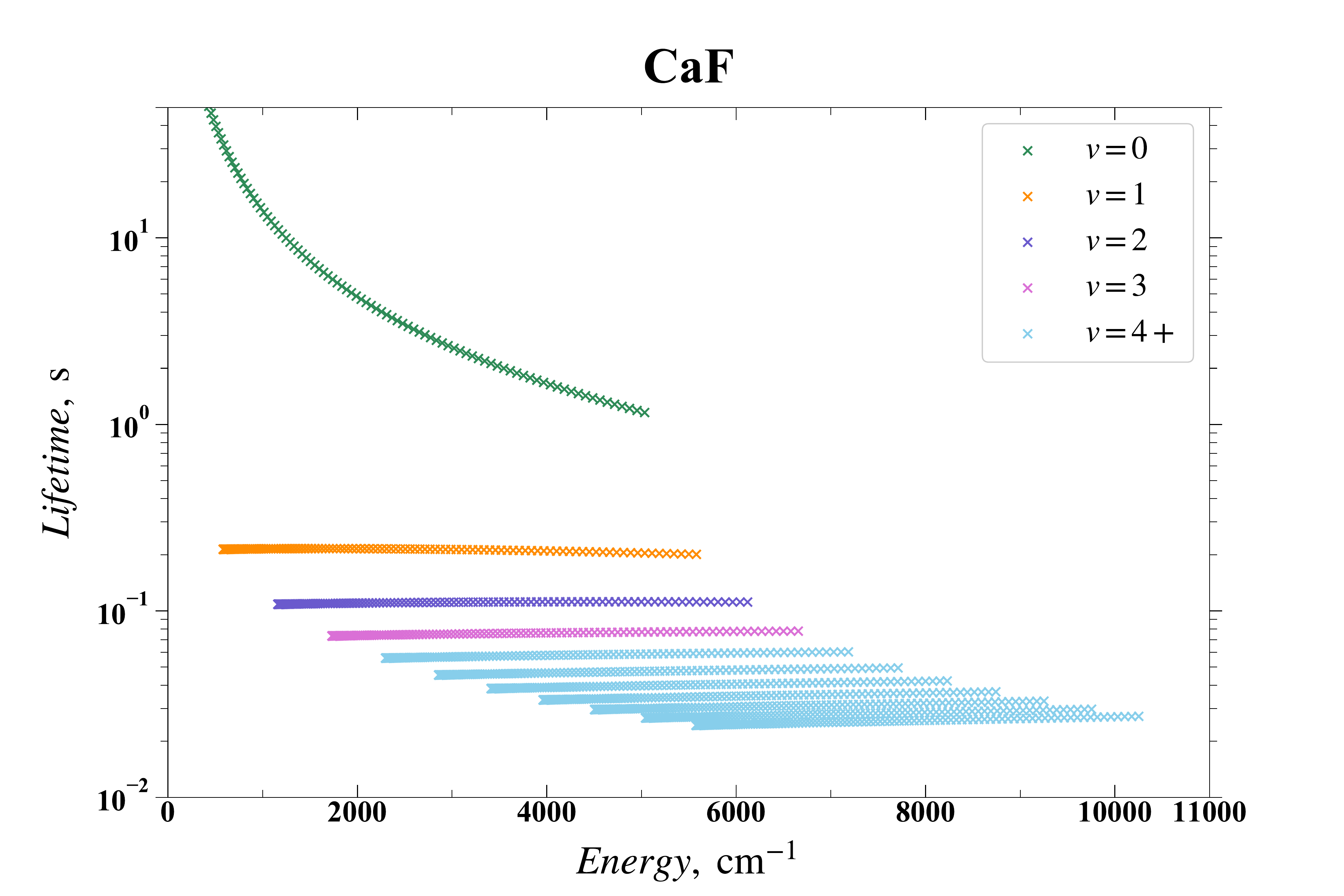}
\includegraphics[width=0.33\textwidth]{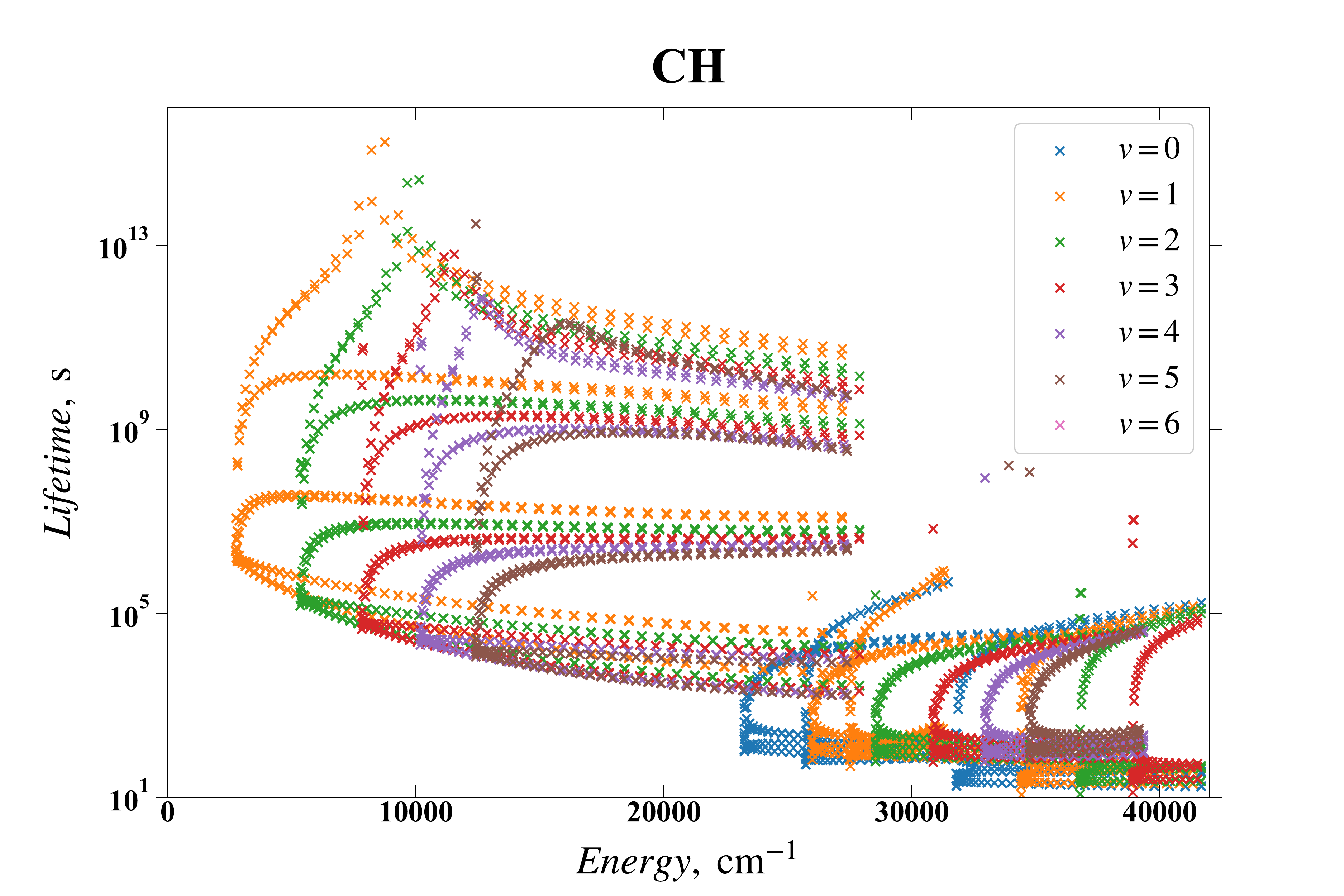}
\includegraphics[width=0.33\textwidth]{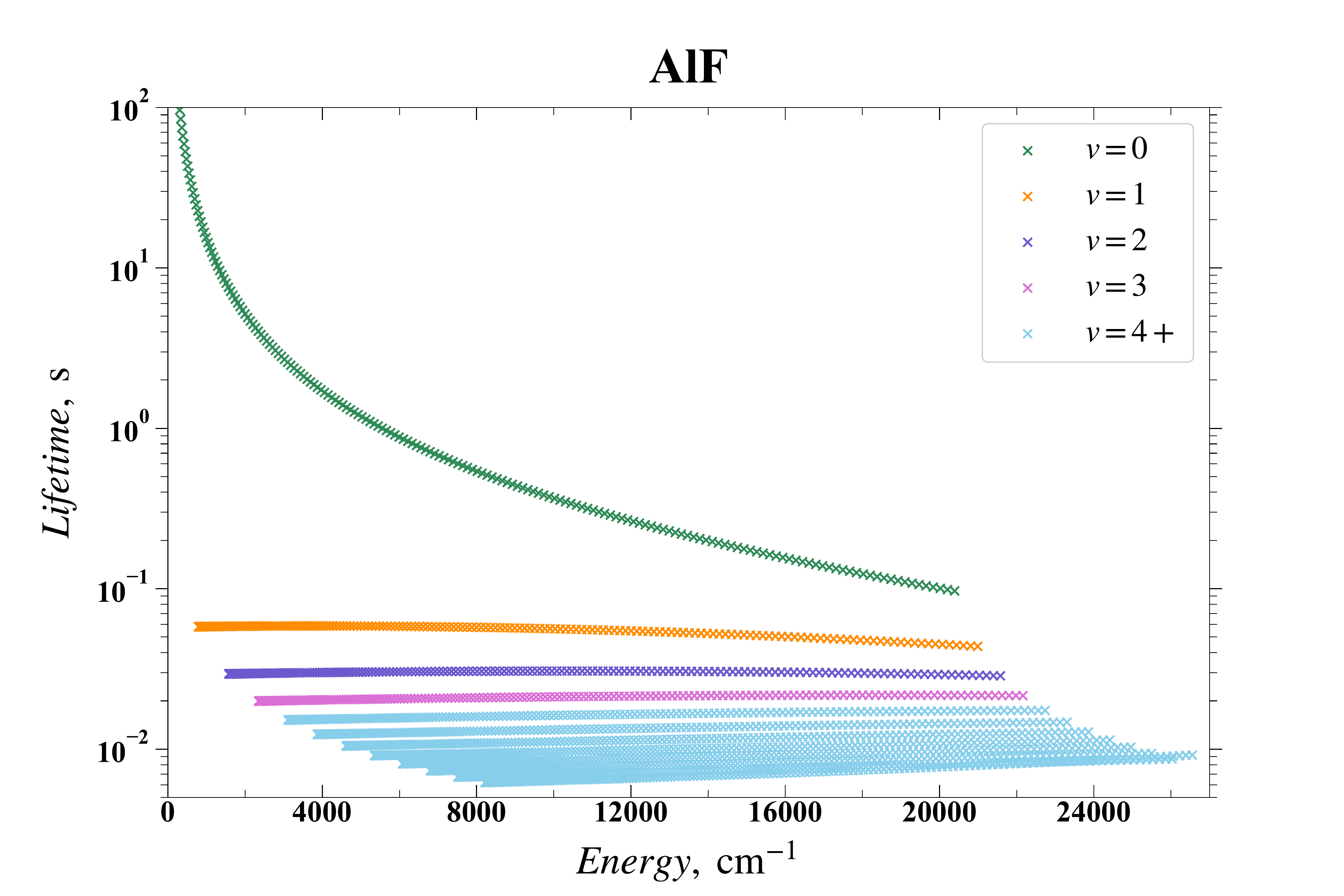}
\includegraphics[width=0.33\textwidth]{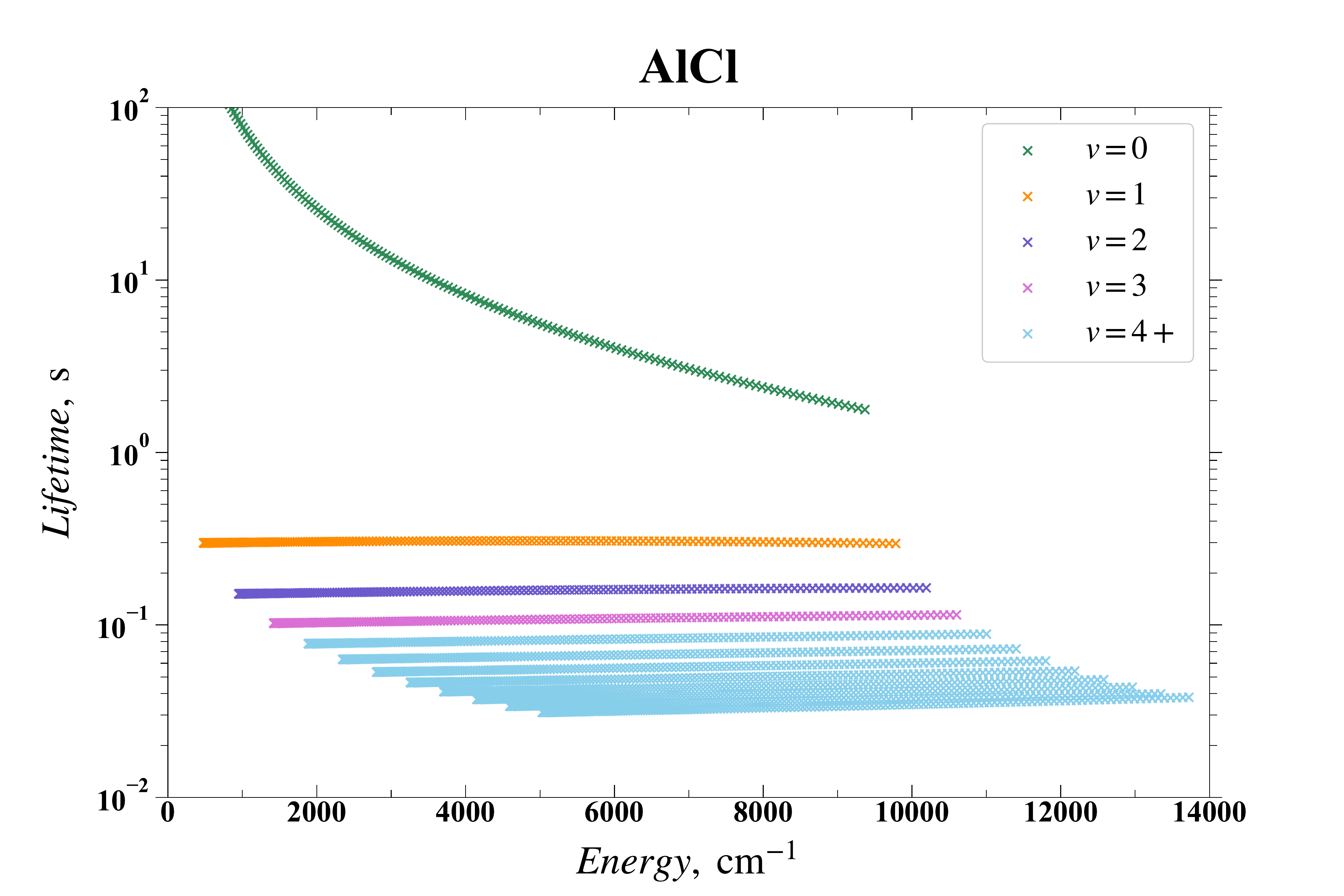}
\includegraphics[width=0.33\textwidth]{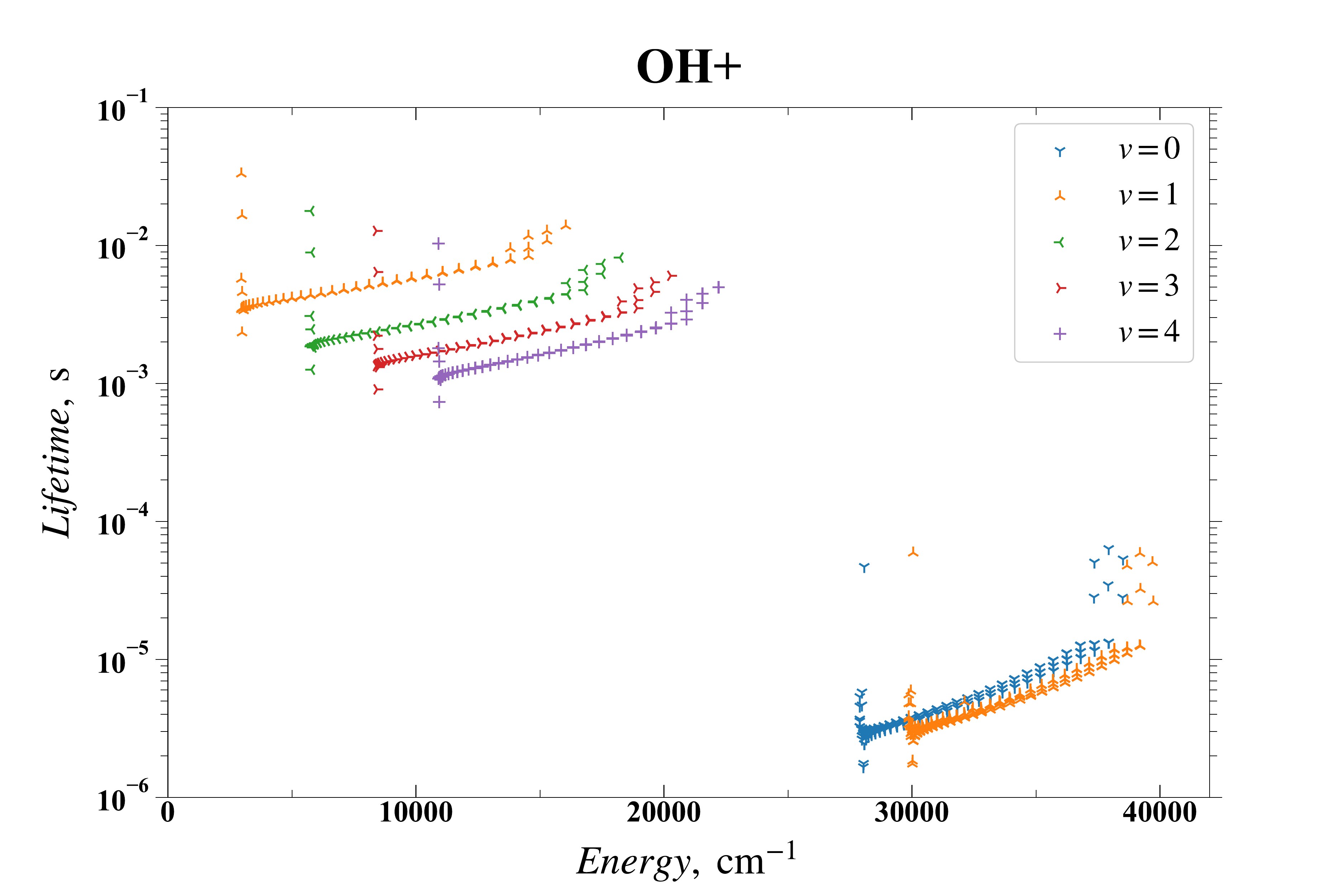}
\caption{Lifetimes as a function of state energy for the line lists considered in this article. The~plots give results for the main isotopologue~only.}
\label{f:lf}
\end{figure}

Lifetimes for each state were generated with ExoCross \citep{jt708} using the methodology of \citet{jt624}. Examples of lifetimes computed using the empirical line lists from Bernath's group  are shown in Figure~\ref{f:lf}.

\section{Conclusions}

The ExoMol database provides molecular line lists valid over extented temperature ranges for many molecules. This paper describes the inclusion in the database of a new set of experimentally generated line lists which were produced externally to the main ExoMol activity. Most of these line lists (all except the one for N$_2$) are from the MoLLIST data set \url{http://bernath.uwaterloo.ca/molecularlists.php} \citep{MOLLIST}.
Inclusion of these line lists significantly improves the coverage of the ExoMol database. They~will
form an important part of the upcoming release of the database \citep{jt799}.


\acknowledgments{Yueqi
\lq\lq Zoe\rq\rq  Na's help with the ExoMol database is greatly appreciated.
We thank Peter F. Bernath for many helpful discussions and for encouraging us to use MoLLIST data;
we thank Colin Western for help with the N$_2$ line list and Geronimo L. Villanueva for helping with line lists for CH, NH, and C$_2$.

This work was supported by the UK Science and Technology Research
Council (STFC) No. ST/R000476/1. Yixin Wang's visit was supported by Physics Boling Class in Nankai University.

}


\end{document}